\documentstyle[psfig]{mn}
\title[Low-mass Hyades binaries]
{Low-mass spectroscopic binaries in the Hyades: a candidate brown dwarf companion}
\author[I.N. Reid \& S. Mahoney]
{I. Neill Reid$^{1}$ and S. Mahoney$^2$\\
$^1$Dept. of Physics \& Astronomy, University of Pennsylvania, 209 S. 33rd
Street, Philadelphia, PA 19104-6396; \\
e-mail: inr@morales.physics.upenn.ed \\
$^2$Hinman Box 1709, Dartmouth College, Hanover, NH 03755 \\}

\begin {document}

\maketitle
 
\begin {abstract}

We have used the HIRES echelle spectrograph in the Keck I telescope to obtain 
high-resolution spectroscopy of 51 late-type M dwarfs in the Hyades cluster. 
Cross-correlating the calibrated data against spectra of white dwarfs allows
us to determine heliocentric velocities with an accuracy of $\pm 0.3$ kms$^{-1}$.
Twenty seven stars were observed at two epochs in 1997; two stars, RHy 42 and RHy 403,
are confirmed spectroscopic binaries. RHY 42 is a double-lined, equal-mass system; 
RHy 403 is a single-lined, short-period binary, P$\sim 1.275$ days.
RHy 403A has an absolute magnitude of M$_I = 10.85$, consistent with a mass of 0.15 M$_\odot$.
The systemic mass function has a value${{(M_2 sin(i))^3} \over {(M_1 + M_2)^2}} = 0.0085$, which,
combined with the non-detection of a secondary peak in the cross-correlation function, implies
$0.095 > M_2 > 0.06s M_\odot$, and the strong possibility that the companion is the
first Hyades brown dwarf to be identified. Unfortunately, the maximum expected angular separation 
in the system is only $\sim 0.25$ mas. \\
Five other low-mass Hyads are identified as possible spectroscopic binaries, based either on
repeat observations or comparison between the observed radial velocity and the value
expected for Hyades cluster members. Combined with HST imaging data, we infer a
binary fraction between 23 and 30\%. \\
All of the stars are chromospherically active. RHy 281 was caught in mid-flare and, 
based on that detection, we estimate a flaring frequency of $\sim 2.5\%$ for
low-mass Hyades stars. Nine stars have rotational velocities, $v \sin(i)$, 
exceeding 20 kms$^{-1}$ and most of the sample have detectable rotation. We
examine the H$\alpha$ emission characteristics of low-mass cluster members and show
that there is no evidence for a correlation with rotation. 

\end{abstract}
 
 
\section {Introduction}

The frequency of binary and multiple systems amongst stars of different masses is
a fundamental parameter in both star formation theory, and understanding mechanisms
for the formation of planetary systems.  
Low-mass stars have received less attention than solar-type G dwarfs, but the available observations
suggest that the multiplicity fraction decreases from $\sim70\%$ or more
at M$\sim 1 M_\odot$ (Duquennoy \& Mayor, 1991) to $\sim35\%$ for M dwarfs (Fisher \&
Marcy, 1991; Reid \& Gizis, 1997a). Most previous surveys have concentrated on field
stars, which span a range of age and abundance. Open clusters offer an alternative
laboratory, where age and abundance are both homogeneous and, within the 
limits of theoretical models, known factors. Cluster analyses, however, must allow
for the possibility that dynamical evolution may have modified key characteristics
since formation. Binaries, for example, are likely to be 
retained preferentially, inflating the apparent binary frequency at a given
spectral type.

The Hyades is the nearest substantial open cluster, and its members have received
considerable observational and theoretial attention. Higher-mass cluster members (spectral
types earlier than mid-K) have
been the subject of both radial velocity observations (Griffin et al, 1988) and
high-resolution infrared speckle interferometry (Patience et al, 1998), as well as
extensive proper-motion surveys capable of detecting wide companions  
(Hanson, 1975; Reid, 1992 and refs within). The
deduced binary frequency is broadly consistent with Duquennoy \& Mayor's analysis of
similar-mass stars in the Solar Neighbourhood. At lower masses, over fifty of the 
cluster M dwarfs were targeted for observation by the Hubble Space Telescope (Gizis \& Reid,
1997; Reid \& Gizis, 1997b), with the Planetary Camera images providing a spatial
resolution comparable to the ground-base speckle data, $\Delta \sim 0.1$ arcsec. Again,
the frequency of companions is consistent with that amongst nearby field dwarfs, although
only six binaries are available for this statistical comparison.

Few accurate radial velocity measurements exist for late-type Hyades dwarfs. Griffin et al
include a few late-K, early-M dwarfs in their survey, and Cochran \& Hatzes (1999)
include early-type M dwarfs in their planet search. Stauffer et al (1997) surveyed
49 early-type M dwarfs, including three non-members, 
identifying seven double-lined spectroscopic binaries 
and one triple system. The faintest star in their sample, vA 127 (RHy 143) has
spectral type M4, and an absolute magnitude M$_V = 12.83$. Data for later type stars
are scarce. Jones et al (1996)  include three of the lowest luminosity cluster members 
(Br 804=RHy 386, Br 816 and Br 262=LP415-20, from Bryja et al, 1993) in their high-resolution 
spectroscopy of low-mass Hyads and Pleiads. Most recently, Terndrup et al (2000) have
observed 18 Hyades dwarfs with M$_V > 11$, including ten in common with the
current sample\footnote{Note that the stars designated Re by Terndrup et al should be
identified using the prefix RHy.}. 

Given the paucity of observations, it is not surprising that few 
spectroscopic binaries have been identified amongst the cluster M dwarfs. Extending 
observations to cover this smaller range of physical separations provides a 
further comparison of the binary frequency relative to that amongst field dwarfs. In
addition, high resolution spectroscopic data provide a means of probing the
physical structure of these stars, notably the range of chromospheric activity and
the distributon of rotatioal velocities.  The current scarcity of observations
reflects the faint apparent magnitudes of these stars, and the consequent difficulties of
obtaining high signal-to-noise data at the appropriate spectral resolution. Those
circumstances have changed with the availability of 8-10 metre class telescopes
equipped with high efficiency echelle spectrographs.  

This paper presents the first results from a survey of late-type Hyades M dwarfs, 
based on observations with the
HIRES spectrograph on the Keck I telescope. 
Most of the targets also have high spatial-resolution HST images.
The following section outlines the observations; section 3 describes the
measurement of radial and rotational velocities; section 4 presents the
techniques used to identify possible binaries; 
section 5 discusses the binaries identified from our observations; section 6
compares rotation and activity in these low mass stars; and section 7
presents our conclusions.

\begin {table*}
\centering
\caption{Hyades cluster members}
\begin{tabular}{lccccclc}\hline
RHy & Other & V & (V-I) & M$_V$ & r (pc.) & Comments \\
\hline\hline
9 & LP 357-86 & 16.73 & 3.18 & 13.45 & 45.3& unresolved HST \\
23 & LP 414-30 & 16.18 & 3.01 & 12.74 &48.8 & unresolved \\
42 & LP 414-794 & 15.30 & 2.93 & 12.04 & 44.9& unresolved \\
46 & LP 414-51 & 16.68 & 3.12 & 13.09 &52.4  & unresolved \\
49AB & LP 414-50 & 15.89 & 3.04 & 12.51 & 47.5& HST binary,  $\Delta = 0".36$, $\delta I ~ 1.6$ \\
60 & LP 414-110 & 16.12 & 3.14 & 12.50 & 52.9 & unresolved \\
64 &  & 16.33 & 3.23 & 13.65 &36.3 & unresolved \\
83 &  & 17.46 & 3.52 & 14.69 & 35.8& unresolved \\
88A & &13.06 & 1.90 & 9.13 & 61.2& binary, $\Delta = 5''$ \\
88B & & 14.11& 2.60 & 10.18& 61.2& \\ 
98 & LP 414-1570 &16.63 & 3.22 & 13.64 &36.8 & unresolved \\
101 & LP 474-1606 & 17.11 & 3.46 & 14.20 &37.9 & unresolved \\
115 & & 16.50 & 3.03 & 12.74 &56.4 & unresolved \\
119AB & LP 414-138 & 17.29 & 3.54 & 14.16 &42.3 & HST binary,  $\Delta = 0".88$, $\delta I ~ 0.1$ \\
126 & LP 414-1895 & 17.33 & 3.38 & 13.55 & 37.8& unresolved \\
129 & LP 474-217 & 15.33 & 2.86 & 12.40 &38.5 & unresolved \\
132 & vA 94 & 15.83 & 3.03 & 12.30 &50.8 & unresolved \\
143 & vA 127 & 16.23 & 3.02 & 12.83 & 37.9& unresolved \\ 
158 & LP475-14 & 16.31 & 3.20 & 13.07 &44.5 & unresolved \\
162 & & 16.54 & 3.21 & 13.63 & 38.2 & unresolved \\
163 & vA 213 & 15.44 & 3.00 & 12.48 & 39.1& unresolved \\
165 & LP 415-19 & 19.67 & 4.34 & 16.41 & 44.9& unresolved \\
182 & LP 4156-35& 15.89 & 3.06 & 12.53 &47.0 & unresolved \\
199 & LP 415-50 & 16.70 & 3.22 & 13.42 &45.3 & unresolved \\
200 & LP 475-59 & 16.07 & 2.89 & 12.56 &50.3 & unresolved \\
202AB & LP 415-51 & 16.70 & 3.11 & 13.34 &37.0 & HST binary, $\Delta = 0".28$, $\delta I ~ 1.8$ \\
219 & vA 352 & 16.37 & 2.87 & 13.03 &46.6 & unresolved \\
221AB & LP 415-71 & 15.88 & 3.10 & 12.54 & 46.7& HST binary,  $\Delta = 0".31$, $\delta I ~ 0$ \\
230 & LP 415-881 & 18.25 & 3.42 & 14.72 &38.7 & unresolved \\
231 & LP 415-875 & 15.89 & 3.11 & 12.49 & 38.6& unresolved \\
240 A & & 18.77 & 3.72 & 15.39 & 47.5& binary,  $\Delta = 2".5$, $\delta I ~ 0$ \\
240 B & & 18.96 & 3.82 & 15.59 & 47.5 & \\
242 & vA 432 & 15.91 & 2.92 & 12.55 &47.0 & unresolved \\
244AB & LP 415-108 & 15.64 & 2.97 & 12.07 & 51.8& HST binary, $\Delta  = 0".14$, $\delta I ~ 1.1$ \\
260 & Ha 430 & 16.40 & 3.07 & 13.52 &37.6 & unresolved \\
281 & LP 415-142 & 18.12 & 3.30 & 14.39 &55.8 & unresolved \\
297 & LP 475-1012 & 17.04 & 3.18 & 13.42 &52.9 & unresolved \\
298 & vA 616 & 15.90 & 2.93 & 12.66 &44.5 & unresolved \\
301 & LP 475-125 & 16.53 & 3.15 & 13.43 &41.6 & unresolved \\
331 & LP 415-206 & 16.43 & 2.99 & 13.30 &42.3 & unresolved \\
346AB & LP 475-176 & 16.03 & 3.18 & 12.84 &43.4 & HST binary,  $\Delta = 0".48$, $\delta I ~ 0.4$ \\
367 &  & 16.45 & 3.13 & 12.94 &50.3 & unresolved \\       
369 & LP 475-1419 & 16.14 & 2.92 & 12.59 &51.3 & unresolved \\
371AB &  LP 415-266 & 15.36 & 3.10 & 12.02 &46.6 & HST binary,  $\Delta = 0".17$, $\delta I ~ 1.3$ \\
376 & LP 415-1692 & 15.80 & 2.90 & 12.52 &45.3 & unresolved \\
377AB & LP 475-214 & 16.24 & 3.15 & 12.65 &52.4 & binary HST,  $\Delta = 1".66$, $\delta I ~ 1.0$ \\
386 & LP 415-2128 & 18.83 & 3.69 & 15.66 &43.0 & unresolved \\
390 & LP 415-1773 & 18.49 & 3.83 & 15.13 &47.0 & unresolved \\
391 &   & 16.42 & 3.35 & 13.08 &46.6 & unresolved \\ 
399 & LP 415-1816 & 15.09 & 2.92 & 12.18 &38.2 & unresolved \\
402 & & 17.77 & 3.25 & 13.94 &58.4 & unresolved \\
403 & LP 415-327 & 18.04 & 3.47 & 14.42 &52.9 & unresolved \\
 & G8-17 & 12.68 & & 9.3& & unresolved \\
\hline
\end{tabular}
\end{table*}

\begin {table*}
\begin{center}
\caption{Field M dwarfs}
\begin{tabular}{cccccclcccc}\hline
Name  & M$_V$ & (V-I) & Spectral type & V$_{hel}$ & References & V$_{hel}$ & MJD \\
      &       &       &               &  kms$^{-1}$ & &kms$^{-1}$ & 50400+\\
\hline\hline
Gl 54.1 & 14.17 & 3.15 & M4.5 &  & L92, RHG & 28.0$\pm0.10$ & 60.20727\\
Gl 83.1 & 14.02 & 3.12 & M4.5 & -29 & L92, RHG, D98 &-28.6$^{ref}$ & 60.20251\\
        &   &   &  &  &  & -28.6$^{ref}$ & 11541.17607 \\
       &   &   &  &  &  & -28.6$^{ref}$ & 11542.18446 \\
Gl 234 A & 13.01 & 3.02 & M4.5 & 12& L92, RHG & 14.85$^{ref}$ & 793.59378\\
       &   &   &  &  &  & 16.2$\pm0.12$ & 11542.46873 \\
LP 771-95 & 11.9 & & M3.5 & & RHG & -1.2$\pm0.13$ & 60.21650\\
LP 771-96A & 12.5 & & M4 & & RHG & -5.5$\pm0.18$  &60.21946\\
Gl 402 & 12.46 & 2.79 & M4 & -1.1 & L92, RHG, MB89 & -1.1$\pm0.20$ & 792.66871\\
Gl 445 & 12.17 & 2.64 & M4 &  & L92, RHG & -111.5$\pm0.25$& 794.67419\\
Gl 447 & 13.50 & 2.98 & M4 & -31.34 & L92, RHG, MB89& -31.2$\pm0.24$& 794.66546\\
       &       &      &    &        &   &    -31.3$\pm0.24$ & 11541.56632 \\
Gl 831AB & 12.50 & 2.99 & M4.5 & -57.1 & L92, RHG, MB89 & -58.75$\pm0.30$ & 793.19816 \\
Gl 876 &  11.82 & 2.74 & M4 & -1.77 & L92, RHG, MB89& -1.95$\pm0.25$& 793.21099\\
 & \\
RHy 80A & &3.17 &M4.5 && R93 & 0.88$\pm0.25$ & 60.37904\\
RHy 80B & &3.2  & M4.5 && R93 & 29.12$\pm0.42$ &  \\
RHy 80A & &3.17 &M4.5 && R93 & 2.35$\pm0.25$ & 792.34271\\
RHy 80B & &3.2 & M4.5 && R93 & 26.4$\pm0.35$ & \\
RHy 110 & &3.03 &M4 && R93 & 60.67$\pm0.25$ & 60.42207 \\
RHy 225 & &2.76 & M3 && R93 & 40.25$\pm0.31$& 792.54615 \\
RHy 254 & &3.22 & M4&& R93 & -4.12$\pm0.20$ & 60.25284 \\
\hline
\end{tabular}
\end{center}
Column 6 identifies the source of the photometry, spectroscopy and radial velocity
measurements listed in columns 2 to 5. \\
L92: photometry from Leggett, 1992 \\
R93: photometry/spectral type from Reid, 1993 \\
RHG: spectral type from Reid et al, 1995a \\
MB89: Radial velocity from Marcy \& Benitz, 1989 \\
D98: Radial velocity from Delfosse et al, 1998
\end{table*}

\section {Observations}

We have obtained high resolution spectroscopy of 53 late-type members of the 
Hyades cluster using the HIRES echelle spectrograph (Vogt et al, 1995) on the 
Keck I telescope. The observations were made in the course of three
allocations: a one night run,
January 13 (UT), 1997; over the three night period December 10 to 12 in the
same year; and on December 29 and 30, 1999. 
Conditions were clear and photometric throughout, with sub-arcsecond seeing. 
All of the  target stars except G8-17 are drawn from Reid's (1993) list of stars with both 
proper motions and photometry consistent with Hyades membership. G 8-17 is
listed as a candidate Hyades member in the Lowell survey (Giclas et al, 1972), and the star has
UBV photomety by Sandage \& Kowal (1986). Our spectrum confirms it as a Hyades member.
The majority of these stars were included 
in Reid \& Gizis' (1997b) HST Planetary Camera imaging programme, which embraced Hyades
members with M$_V > 11.9$ (spectral type later than M3). Several
are therefore known to be spatially-resolved binary systems. Table 1 lists photometric
data, from Reid (1993) and Leggett et al (1994), for stars in the present
sample. Individual distances are estimated from the proper motions using the
convergent point method and the results from Perryman et al's (1998) Hipparcos
analysis (see section 4). These values are typically 5\% smaller than 
distances cited in Reid (1993).

In addition to the Hyades members, we obtained observations of
field M dwarfs of comparable spectral types, as well
as several candidate Hyads which prove to be non-members.
Photometry and spectral types for those stars are listed in Table 2, 
together with heliocentric velocity measurements, both from the published literature
(Marcy \& Benitz, 1989; Delfosse et al, 1998) and from our own observations. 
One of the Hyades non-members, RHy 80, is a double-lined spectroscopic
binary, Table 2 lists velocity measurements at two epochs.

Our observations were made using a 0".86 second slit, which gives
a resolution of 45,000, or 6.67 km s$^{-1}$. With an echelle angle of -0$^o$18
and a cross-disperser angle of 1$^o$37, the spectra provide
partial coverage of the wavelength region from 6350 \AA\ and 8730\AA\ in 16 orders,
including H$\alpha$ and the atmospheric A and B bands. We adopted a standard exposure time
of 600 seconds for stars with 13$<$V$<$17, increasing to integration time to between 900 and
1800 seconds for fainter stars. 

The images were flat-field
calibrated and bias subtracted in the usual manner, and the spectral orders
extracted using software written by T. Barlow. The wavelength calibration was
determined from thorium-argon lamp exposures made at the start of each night,
using standard IRAF routines to apply that calibration to the individual observations.
Previous experience has shown that this technique provides velocities accurate to 
$\pm1$ kms$^{-1}$. Interspersing calibration arcs between programme observations does not
necessarily lead to improve the stability of the wavelength scale. However, other means are 
available to establish the velocity zeropoint with higher accuracy, as described in the
following section. We have not attempted to set any of the spectra on an absolute flux scale.

\section {Radial velocity and stellar rotation measurements}

The primary aim of this programme is to identify spectroscopic binaries in the
Hyades cluster. In general, achieving such a goal demands multiple, high-accuracy
spectroscopic observations: our initial reductions provide velocities accurate to
only $\pm 1$ kms$^{-1}$, and we have only one observation of almost half of the
stars listed in Table 1. As discussed further below,  we have been able to
refine the accuracy of our velocity determinations.
Moreover, we can utilise even single observations for binary detection, since
we have the advantage of targetting stars with known space motion. 

The standard method of determining radial velocities is cross-correlation of the
calibrated spectra against data for a standard star of similar spectral type and
known radial velocity (Tonry \& Davis, 1979). Uncertainties in the derived 
velocities depend both on random errors, which can be estimated from the 
width of the cross-correlation peak and inter-order agreement in the
derived velocities, and on the accuracy of the wavelength calibration. In
past observations, we
have noted systematic differences of 1 to 1.5 kms$^{-1}$ in repeated HIRES obervations of
individual stars. These offsets are present in both sky and stellar spectra, even if calibrated
using adjacent Th-Ar arcs. The discrepancies may originate either
through differences in the illumination of the slit, or  
internal variations in the optical elements within the HIRES instrument.

Given those concerns, we adopt a two stage approach to determining heliocentric
velocities. First, the observations are cross correlated against standard
stars in the usual manner; second, we use terrestial features in
the stellar spectra to set the observations on a uniform zeropoint. 

\subsection {Relative velocity determination}

In cross correlation, the spectral type of the standard template must be
a good match to that of the programme stars. For the current observations, 
the two field stars which best satisfy this constraint are the M4.5 dwarfs 
Gl 83.1 (January, 1997 and December, 1999) and Gl 234A (December, 1997).  
Neither has published high-precision velocity measurements:
heliocentric velocities for both systems have been measured by Delfosse et al (1998),
but the results are cited only to the nearest kms$^{-1}$. Moreover, Gl 234A is  
the primary component in a wide binary, with a $\sim 4.17$ AU, P $\sim 16.60$
years,  ${M_2 \over M_1} \sim 0.5$ (Henry \& McCarthy, 1993). 

We have therefore determined heliocentric velocities for both of our template stars
by cross-correlation against our data for field M dwarfs observed
by Marcy \& Benitz (1989): Gl 402, 447 and 876. Marcy \& Benitz also observed
Gl 831, but at only one epoch. Since that star is known to be a binary,
it is not included in our zeropoint determination. Table 2 lists the results
for all of the field stars. 
We derive a heliocentric velocity of -28.6 kms$^{-1}$ for Gl 83.1, with no
significant variation evident in our three observations. In contrast, Gl 234A has
a heliocentric velocity of 14.85 kms$^{-1}$ in December 1997, while we measure
16.17 kms$^{-1}$ in December, 1999. 

The individual spectra were cross-correlated, order by order, against the
appropriate template using the IRAF routine {\sl fxcor}. The velocity differences
computed by the latter routine 
includes heliocentric corrections.  Three echelle orders are dominated by terrestial
features: 6800-6890\AA, the B band; 7520-7630\AA, the A-band; and 8220-8330\AA.
We excluded those orders when determining the mean velocity measured for
the programme stars, 
In most cases, the remaining orders correlate well, with cross-correlation
peaks of amplitude 0.7-0.9 and formal uncertainties of
0.3 to 0.7 kms$^{-1}$ in the individual velocity measurements. We have
averaged the latter measurements to derive the radial velocities, $\langle V \rangle$
listed in Tables 3 and 4, citing the rms dispersion about the mean, $\sigma_V$, as an 
estimate of the uncertainty. In most cases,
the latter values are 0.15 to 0.3 kms$^{-1}$.  Approximately 20\% of the sample have
substantial rotational velocities, $v \sin(i) > 15$kms$^{-1}$. 
These stars have higher uncertainties in the derived velocities. 

\begin {table*}
\centering
\caption{Radial velocity measurements}
\begin{tabular}{lcccclcccc}\hline
Star & $\langle V \rangle$ & $\sigma_V$ & $\Delta$V$_{WD}$ & V$_{hel}$
& Star & $\langle V \rangle$ & $\sigma_V$ & $\Delta$V$_{WD}$ & V$_{hel}$ \\
   &  kms$^{-1}$ & kms$^{-1}$ & kms$^{-1}$ & kms$^{-1}$ & &  kms$^{-1}$ &
 kms$^{-1}$ & kms$^{-1}$& kms$^{-1}$ \\
\hline\hline
 & & January, 1997 &  field stars \\
Gl 83.1 & ref & & 0.57 & -28.6 & Gl 54.1 & 28.17 & 0.18 & 0.76 & 27.98 \\
LP 771-95 & -1.28 & 0.24 & 0.52 & -1.23 & LP 771-96AB & -5.67 & 0.30 & 0.38 & -5.48 \\
RHy 80 & 1.67 & 0.45 & -0.11 & 2.35 & RHy 110 & 59.26 & 0.24 & -0.84 & 60.67 \\
RHy 254 & -4.49 & 0.40 & 0.20 & -4.12 \\
 & & January, 1997 &  Hyads \\
RHy 9 & 36.82 &  0.28 & 0.41 & 36.98 & RHy 23 & 36.19 & 0.25 & 0.33 & 36.43 \\
RHy 42A & 30.28 & 0.74 & 0.09 & 30.76 & RHy 46 & 36.45 & 0.33 & 0.05 & 36.97 \\
RHy 42B & 41.70 & 0.81 & 0.09 & 42.18 \\
RHy 49 & 36.41 & 3.40 & -0.25 & 37.22 & RHy 60 & 36.33 & 0.78 & -0.10 & 37.00 \\
RHy 64 & 35.71 & 0.86 & -0.23 & 36.51 & RHy 88A & 38.50 & 0.27 & 0.00 & 39.07 \\
RHy 88B & 33.82 & 0.19 & -0.15 & 35.04 & RHy 98 & 38.13 & 3.56 & -0.48 & 39.19 \\
RHy 101 & 37.37 & 0.98 & -0.28 & 38.22 & RHy 115 & 38.21 & 0.28 & -0.13 & 38.91 \\
RHy 119 & 36.70 & 0.41 & -0.32 & 37.59 & RHy 126 & 36.18 & 2.61 & -0.54 & 37.29 \\
RHy 129 & 38.12 & 0.20 & -0.37 & 39.06 & RHy 143 & 37.62 & 0.28 & -0.43 & 38.62 \\
RHy 158 & 38.77 & 0.38 & -0.51 & 39.85 & RHy 163 & 36.70 & 7.41 & -0.82 & 38.09 \\
RHy 244 & 38.13 & 0.23 & 0.74 & 37.96 & RHy 260 & 40.32 & 0.24 & 0.30 & 40.59 \\
RHy 297 & 39.83 & 0.74 & 0.11 & 40.29 & RHy 298 & 40.25 & 0.26 & 0.19 & 40.43 \\
RHy 301& 39.73 & 0.95 & -0.92 & 40.98 & RHy 309 & 39.33 & 0.24 & -0.92 & 40.82 \\
 & & December, 1997 & field stars \\
Gl 402 & -4.68 & 0.25 & -2.81 & -1.10 & Gl 445 & -113.28 & 0.26 & -1.01 & -111.50 \\
Gl 447 & -33.55 & 0.24 & -1.60 & -31.18 & Gl 876 & -3.74 & 0.20 & -1.13 & -1.97 \\
RHy 110 & 56.11 & 0.09 & -3.24 & 60.12 & Gl 234A &ref & & 0.77 & 14.85\\
 & &  December, 1997 & Hyads \\
RHy 9 & 33.03 & 0.25 & -3.37 & 37.17 & RHy 23 & 32.35 & 0.29 & -3.33 & 36.45 \\
RHy 42A & 14.96 & 1.01 & -3.15 & 18.88 & RHy 46 & 32.68 & 0.36 & -3.46 & 36.91 \\
RHy 42B & 53.67 & 0.53 & -3.15 & 57.59 \\ 
RHy 60 & 32.31 & 0.78 & -3.61 & 36.69 & RHy 64 & 31.83 & 0.87 & -3.49 & 36.09 \\
RHy 83 & 33.27 & 0.45 & -3.62 & 37.66 & RHy 88A & 35.05 & 0.35 & -3.30 & 39.13 \\
RHy 88B & 32.18 & 0.24 & -3.18 & 36.13 & RHy 98 & 32.73 & 1.65 & -3.29 & 36.89 \\
RHy 101 & 33.32 & 0.41 & -3.65 & 37.74 & RHy 115 & 34.66 & 0.39 & -3.63 & 39.06 \\
RHy 119 & 33.91 & 0.37 & -3.01 & 37.69 & RHy 129 & 34.96 & 0.39 & -3.19 & 38.92 \\
RHy 143 & 34.43 & 0.40 & -3.31 & 38.51 & RHy 158 & 33.55 & 0.37 & -3.31 & 37.63 \\
RHy 162 & 34.28 & 2.48 & -2.84 & 37.89 & RHy 163 & 34.82 & 0.70 & -2.87 & 38.46 \\
RHy 182 & 35.52 & 0.39 & -2.98 & 39.27 & RHy 199 & 35.20 & 0.71 & -2.84 & 38.81 \\
RHy 200 & 36.27 & 0.52 & -2.79 & 39.83 & RHy 202 & 35.19 & 0.38 & -2.75 & 38.71 \\
RHy 219 & 35.87 & 0.54 & -2.63 & 39.27 & RHy 221 & 37.08 & 0.33 & -2.70 & 40.55 \\
RHy 225 & 36.91 & 0.31 & -2.57 & 40.25 & RHy 231 & 35.68 & 0.42 & -2.65 & 39.10 \\
RHy 230 & 37.40 & 1.38 & -1.51 & 39.68 & RHy 240A & 37.59 & 1.34 & -1.37 & 39.73 \\
RHy 240B & 36.22 & 0.21 & -1.24 & 38.33 & RHy 260 & 38.73 & 0.25 & -1.08 & 40.68 \\
RHy 281 & 31.26 & 1.32 & -1.29 & 33.32 & RHy 297 & 38.42 & 0.54 & -1.14 & 40.33 \\
RHy 301 & 38.27 & 0.54 & -1.04 & 40.08 & RHy 331 & 38.23 & 0.30 & -1.13 & 40.13 \\
RHy 346 & 39.00 & 0.20 & -1.10 & 40.90 & RHy 367 & 39.18 & 0.20 & -1.15 & 41.10 \\
RHy 371 & 36.37 & 0.67 & -1.09 & 38.23 & RHy 377 & 40.65 & 0.47 & -0.98 & 42.40 \\
G 8-17 & 34.96 & 0.22 & -1.12 & 36.85 & RHy 386 & 39.41 & 0.51 & -1.10 & 41.30 \\
RHy 391 & 37.88 & 0.82 & -1.04 & 39.69 & RHy 399 & 38.97 & 0.25 & -1.02 & 40.76 \\
RHy 402 & 39.23 & 0.52 & -0.92 & 40.92 & RHy 403 & 41.22 & 0.47 & -0.97 & 42.96 \\
RHy 165 & 36.82 & 2.10 & -0.92 & 38.51 & RHy 390 & 39.12 & 1.63 & -1.05 & 40.94 \\
RHy 242 & 39.22 & 0.41 & -0.46 & 40.45 & RHy 132 & 37.56 & 0.41 & -0.50 & 38.83 \\
RHy 376 & 39.90 & 0.32 & -0.33 & 41.00 & RHy 83 & 36.79 & 0.52 & -0.28 & 37.84 \\
RHy 46 & 35.58 & 0.26 & -0.54 & 36.89 & RHy 46 & 35.61 & 0.29 & -0.35 & 36.73 \\
RHy 46 & 35.73 & 0.32 & -0.41 & 36.91 & RHy 230 & 38.46 & 1.59 & -0.57 & 39.80 \\
RHy 281 & 30.97 & 1.07 & -1.06 & 32.80 & RHy 297 & 38.42 & 0.39 & -1.22 & 40.41 \\
RHy 386 & 39.60 & 0.79 & -1.21 & 41.58 & RHy 402 & 38.95 & 0.78 & -1.25 & 40.97 \\
RHy 403 & 71.41 & 0.74 & -1.58 & 73.76 \\
\hline
\end{tabular}
\end{table*}

\setcounter{table}{2}
\begin {table*}
\centering
\caption{Radial velocity measurements}
\begin{tabular}{lcccclcccc}\hline
Star & $\langle V \rangle$ & $\sigma_V$ & $\Delta$V$_{WD}$ & V$_{hel}$
& Star & $\langle V \rangle$ & $\sigma_V$ & $\Delta$V$_{WD}$ & V$_{hel}$ \\
   &  kms$^{-1}$ & kms$^{-1}$ & kms$^{-1}$ & kms$^{-1}$ & &  kms$^{-1}$ &
 kms$^{-1}$ & kms$^{-1}$& kms$^{-1}$ \\
\hline\hline
& & December, 1999 & field stars \\
Gl 83.1 & ref &  & 0.71 & -28.6 & Gl 447 & -30.63 & 0.32 & 1.57 & -31.49 \\
Gl 83.1 & ref & & 0.0 & -28.6 & Gl 234A & 16.35 & 0.12 & 0.18 & 16.17 \\
Gl 447 & -30.90 & 0.15 & -31.21 & 0.31 \\
& & December, 1999 & Hyads \\
  RHy 42 & 36.31 & 0.16 & 0.95 & 36.07 & RHy 281 & 32.37 & 1.19 & 0.42 & 32.66 \\
RHy 403 & 69.01 & 0.56 & 0.29 & 69.42 & RHy 42 & 35.88 & 0.25 & 0.0 & 36.59 \\
RHy 403 & 74.68 & 0.42 & -0.02 & 75.41 & RHy 42 & 36.18 & 0.33 & -0.15 & 37.04 \\
RHy 403 & 78.02 & 0.34 & -0.10 & 78.83 & RHy 42 & 36.23 & 0.28 & 0.48 & 33.19 \\
RHy 281 & 32.36 & 0.51 & -0.21 & 33.19 & RHy 403 & 79.24 & -0.04 & 79.99 \\
RHy 42 & 36.10 & 0.31 & 0.42 & 36.39 & RHy 403 & 78.49 & 0.40 & 0.74 & 78.46 \\
RHy 42 & 36.74 & 0.27 & 0.44 & 37.01 & RHy 403 & 75.24 & 0.64 & 0.80 & 75.15 \\
RHy 42 & 37.08 & 0.31 & 0.72 & 37.07 & RHy 403 & 72.48 & 0.44 & 1.03 & 72.16 \\
RHy 342 & 36.79 & 0.16 & 0.04 & 36.75 & RHy 403 & 20.27 & 0.46 & 0.04 & 20.23 \\
RHy 42 & 35.66 & 0.57 & 0.07 & 35.59 & RHy  403 &27.72 & 0.64 & 0.03 & 27.69 \\
RHy 42 & 36.39 & 0.11 & -0.46 & 36.85 & RHy 403 & 36.80 & 0.37 & -0.40 & 37.20 \\
RHy 42 & 36.24 & 0.17 & -0.35 & 36.59 & RHy 281 & 32.02 & 1.19 & -0.75 & 32.77 \\
RHy 403 & 46.33 & 0.57 & -0.71 & 47.04 & RHy 42 & 35.92 & 0.28 & -0.43 & 36.35 \\
RHy 403 & 56.50 & 0.35 & -0.23 & 56.73 & RHy 42 & 36.17 & 0.35 & -0.42 & 36.59 \\
RHy 403 & 63.94 & 0.33 & -0.19 & 64.13 \\
\hline
\end{tabular}
\end{table*}

\subsection {Establishing the velocity zeropoint}

The optimum method of wavelength calibration is to obtain a calibration spectrum simultaneously
with the programme star data. Current surveys for extrasolar planets, which require
accuracies exceeding 10 ms$^{-1}$, achieve this goal by inserting an iodide cell
below the slit, superimposing iodine absorption lines on
the stellar spectrum (Marcy \& Butler, 1992). Unfortunately, the Hyades M dwarfs are
too faint to permit use of this technique. However, Earth's atmosphere provides a
ready made reference frame, superimposing a grid of O$_2$ and OH 
lines on each stellar spectrum. Observations have shown that the
wavelengths of these lines are stable to $\sim40$ ms$^{-1}$, insufficient for
extrasolar-planet searches, but more than adequate for current purposes. 

The terrestial atmosphere supplies both emission and absorption lines. Of these
two options, the latter is preferable for calibration purposes, since, as with the
iodide cell, the optical path followed by the calibration spectrum is, perforce, identical
to the stellar spectrum. In principle, we could estimate the drift in
zeropoint using  the Hyades/Gl 83.1/Gl 234A cross-correlation results
for orders dominated by terrestial absorption. However, stellar features 
weaken the correlation.  We therefore use white dwarf stars as our 
terrestial templates.

G74-7, our January 1997 template, is a DAZ white dwarf, with strong Ca II H \& K absorption;
HS 0507A, used for our December 1997 observations,  is the brighter of a pair
of DA white dwarfs detected in the course of the Hamburg Schmidt Quasar survey (Hagen
et al, 1995); and
40 Eridani B, December 1999,  is the brightest accessible DA white dwarf.
With the exception of broad H$\alpha$ lines, the intrinsic spectrum is a featureless
continuum at red wavelengths; all absorption features are due to the
Earth's atmosphere. These stars provide excellent templates for calibrating
zeropoint drift of the wavelength scale.
\begin{figure}
\psfig{figure=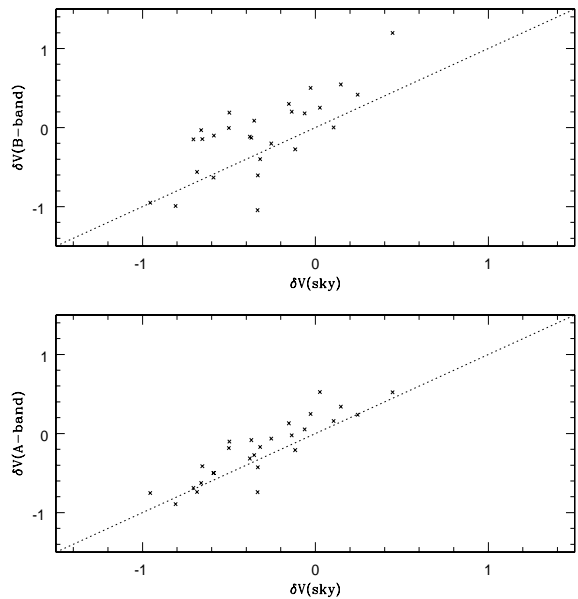,height=8cm,width=8cm
,bbllx=8mm,bblly=57mm,bburx=205mm,bbury=245mm,rheight=10cm}
\caption{Comparison between velocity offsets for individual spectra derived by 
cross-correlation using atmospheric emission lines (abscissa) and the A- and B-band 
absorption features. Data are plotted from January, 1997, using G74-7 as
the white dwarf template. }
\end{figure}

Cross-correlating the programme stars against white dwarf spectra 
produces strong correlations in the orders which include the O$_2$ A- and B-bands.
Figure 1 compares the derived offsets from our January observations against  
the drift in velocity measured using the emission-line sky spectrum from the
same observations. The agreement between the two independent measurements is
good, with
\begin{displaymath}
V_{B-band} - V_{em} = 0.21 \pm 0.34 \ {\rm km s^{-1}}
\end{displaymath}
and
\begin{displaymath}
V_{A-band} - V_{em} = 0.10 \pm 0.19 \ {\rm km s^{-1}}
\end{displaymath}
\begin {table*}
\centering
\caption{Comparison with previous observations: Hyades stars}
\begin{tabular}{lccccl}\hline
Star & $V_{hel}$ & $v \sin(i)$ &  $V_{hel}$ & $v \sin(i)$ & reference \\
     & this paper & & other & & \\
\hline\hline
RHy 143 & 38.6 & 4 & 37.5 & 6.5 & Stauffer et al, 1997 \\
RHy 225 & 40.3 & $< 2.5$ & 39.2 & $<6$ & \\
RHy 386 & 41.5 & 9 & 41.1 & 11 & Jones et al, 1996 \\
RHy 23 & 36.4 & 4 & 36.3 & 6.5 & Terndrup et al, 2000 \\
RHy 60 & 36.85 & 19 & 37 & 19 &\\
RHy 115 & 38.9 & 7 & 39.0 & 9 \\
RHy 119 & 37.6 & 9 & 37.1 & 9 \\
RHy 158 & 37.6/39.5 & 11 & 40.9 & 11 \\
RHy 202 & 38.7 & 2.5 & 38.8 & $<5$ \\
RHy 240b & 38.2 & 14.5 & 40.5 & 14.5 \\
RHy 346 & 40.9 & 3.5 & 41.1 & 6 \\
RHy 386 & 41.5 & 9 & 41.9 & 11.5 \\
RHy 390 & 40.9 & 18 & 40.5 & 18 \\
\hline
\end{tabular}
\end{table*}
The amplitude of variation, $\pm 1$ kms$^{-1}$,  is typical for a single night. 
Given these results, we have combined the A-band and B-band measurements, giving the former
twice the weight of the latter, to derive the appropriate correction. Those values
are listed in Table 3 as $\Delta$V$_{WD}$, and the corrected, heliocentric velocities are listed as
V$_{hel}$. 

Table 4 compares our heliocentric velocity determinations against previously
published results. Setting aside RHy 158, for reasons discussed in the following section, the
mean residual, in the sense of the current measurements {\sl minus} literature data, is
\begin{displaymath}
\Delta V_{R-O} \ = \ 0.04 \pm 0.88 \ {\rm km \ s^{-1}}, {\rm 12 \ stars}
\end{displaymath}

\subsection {Rotation}
 
Tonry \& Davis (1979) originally demonstarted that the
width of the peak in the cross-correlation function is dependent on the line profiles
of both template and programme object. In extragalactic astronomy, this property
is used to measure velocity dispersion; in stellar astronomy, rotation is
usually the dominant contributor to line profile broadening. 
We have used this technique to estimate rotational velocities for the stars
in our sample.

The relation between the measured width of the cross-correlation peak,
fitted either with a Gaussian or parabolic profile, and $v \sin(i)$, the
projected stellar rotation, is determined using spectra of a 
slowly rotating star which have been broadened artificially to match known rotational
velocities. Cross-correlating those spectra against a template, also
with known low rotation and of similar spectral type, provides the necessary calibration. We have used
our observations of Gl 447 (rotation star) and Gl 402 (template) to
calibrate our data. Both stars have negligible rotation, 
$v \sin(1) < 2.0$ kms$^{-1}$ and $< 2.3$ kms$^{-1}$ respectively (Delfosse et al, 1998),
and both are well matched in spectral type to the Hyades programme stars. 

We chose two orders, $\lambda \lambda 7360 -  7460$\AA\ and $\lambda \lambda 7840 - 7960$\AA,
for these measurements, avoiding both TiO bandheads and the stronger terrestial
features. The Gl 447 spectrum was broadened following Gray's (1976) prescription.
The HIRES data have a resolution of R$\sim 45,000$, 
corresponding to a velocity of $v_{FWHM} = 6.67$ kms$^{-1}$. This sets a lower detection
limit of $v \sin(i) \sim 2.5$ kms$^{-1}$ for our rotation measurements. 
Results for the full sample of Hyades stars are tabulated and discussed in section 5.
Table 4 includes comparison of  our measurements against previously-published
results. We derive slightly lower rotational velocities for the slow rotators.
However, the agreement is excellent for moderately and rapidly rotation stars.

\section {Identifying spectroscopic binaries}

The conventional method of identifying spectroscopic binaries is to compare
velocity measurements obtained at different epochs. Table 5 lists the
relevant data from our observations, including both V$_{hel}$ and the Modified Julian 
Date for each observaton. Almost half of the sample have only
a single observation, but even those observations
can be used to identify candidate spectroscopic binaries since 
our targets are members of the Hyades cluster.

Perryman et al (1998) have used astrometric data from
the Hipparcos satellite and CORAVEL radial velocity observations to determine
the mean distance of Hyades cluster members (46.34$\pm0.27$ pc.) 
and the cluster space motion (45.72 kms$^{-1}$). As noted above, those
parameters correspond to a distance scale shorter by $\sim$5\%  than that adopted 
in Reid (1993).  In traditional  open cluster analyses, 
absolute proper motion measurements are used to determine the location in ($\alpha, \delta$)
of the convergent point. Given a known space motion for the cluster, the radial velocity
of an idealised cluster member, V$^P_{rad}$, depends only on the angular distance from the convergent
point, $\lambda_{CP}$. Similarly, the proper motion of a cluster members depends on both $\lambda_{CP}$
and distance, allowing distance estimates to individual cluster members (Table 1).

Circumstances are more complicated for a relatively old cluster, such as the Hyades.
As Perryman et al point out, the velocity dispersion of
$\sigma \sim 0.2$ kms$^{-1}$  leads to an intrinsic fuzziness
in the convergent `point'. Thus, when we invert the process, and predict V$_{rad}$ based
on an idealised convergent point, those predictions are uncertain by $\pm0.2$ kms$^{-1}$.
Indeed, since the Hyades is sufficiently old to have undergone significant mass segregation,
the velocity dispersion (and corresponding uncertainty in V$^P_{rad}$) may
be as high as 0.3-0.4 kms$^{-1}$ for low-mass stars. 
Nonetheless, taking into account these uncertainties,
a significant difference between the observed velocity and the expected velocity
indicates either binary motion, or non-membership of the cluster. In the case of Hyades M dwarfs,
chromospheric activity provides a means of deciding between those alternatives.

We have adopted the convergent point determined by Perryman et al from their
maximum likelihood analysis: they derive ($\alpha=96.5$, $\delta=5.8$) from
a sample of 148 Hyades members. We also use 
their estimate of the Hyades space motion, V = 45.72 kms$^{-1}$. Table 5 
lists both the expected radial velocity, $V^P$, for each Hyades star, and
the observed-computed residual, 
\begin{displaymath}
\delta V^P_{hel} \ = \ V_{hel} - V^P 
\end{displaymath}
We note that the majority of the residuals are positive, with
$\langle \delta V^P_{hel} \rangle = 0.64 \pm 0.68$ (excluding candidate
binaries),
suggesting that the kinematic model we have adopted for the cluster motion may require 
revision. The agreement is adequate for current purposes.

In order to assess the significance of differences
between the predicted and observed velocities, we need to determine the uncertainty, 
$\epsilon_V$, associated with $\delta V^P_{hel}$. These uncertainties stem from
two sources: empirical measuring accuracy, and intrinsic dispersion in cluster motions.   
Our data provide several means of estimating the former:
\begin{enumerate}
\item five observations in 1997 of RHy 46 give $\langle V_{hel} \rangle = 36.88 \pm 0.09$ kms$^{-1}$;

\item thirteen observations (December, 1999) of RHy 42 give $\langle V_{hel} \rangle = 36.57 \pm 0.42$ kms$^{-1}$;

\item five observations (1997-1999) of RHy 281 give $\langle V_{hel} \rangle = 32.95 \pm 0.29$ kms$^{-1}$

\item twenty-one repeat observations of twenty stars with $| \Delta V_{epoch}| < 1$ kms$^{-1}$
give a mean difference of \\ -0.115$\pm0.34$ kms$^{-1}$. 
\end{enumerate}
On that basis, we estimate that an rms uncertainty of 0.35 kms$^{-1}$ is appropriate
for the heliocentric velocity measurements given in Table 3. 
If we assume that the intrinsic velocity
dispersion of Hyades M dwarfs has a similar value, then we derive a minimum value of
$\epsilon_V = 0.5$kms$^{-1}$. If $\sigma_V$, the order-to-order rms dispersion of 
velocity measurements for a given observation, exceeds 0.5 kms$^{-1}$ (Table 3), then 
we let $\epsilon_V = \sigma_V$. Column 7 in Table 4 gives the ratio between $\delta V^P_{hel}$
and $\epsilon_V$ for each star.

\begin {table*}
\centering
\caption{Heliocentric velocities}
\begin{tabular}{lcccccccc}\hline
RHy & MJD & V$_{hel}$ & $\epsilon_V$ & V$^P$ & $\delta V^P_{hel}$ & $\delta V / \epsilon_V$ & $\Delta V_{epoch}$& H$\alpha$ EW\\
  & 50000+& kms$^{-1}$& kms$^{-1}$& kms$^{-1}$ & kms$^{-1}$& &kms$^{-1}$ & \AA\\
\hline\hline
9 & 460.31252 & 36.98 & 0.5 & 36.12 & 0.86 & 1.7 & &5.1\\
 & 792.24254 & 37.22 & 0.5 &  & 1.1 & 2.2 & 0.24 & 3.8\\
23 & 460.32099 & 36.42 & 0.5 & 35.70 & 0.72 & 1.4 & & 3.4\\
  & 792.25098 & 36.50 & 0.5 &  & 0.8 & 1.6 & 0.08 & 4.4\\
42 & 460.32924 &A 30.77 & 0.7 & 36.19 &  &  & &3.4\\
   &           &B 42.18 & 0.8 &  &0.28 & 0.4 &\\
   & 792.25906 &A 18.88 & 0.6 &  &  &  & & 4.3 \\
   &          &B 57.59 & 0.5 &  & 2.0 & 4 \\ 
   &1541.18103 & 36.07 & 0.5 &  & -0.12 & -0.2 & & 3.5 \\
   & 1541.23980 & 36.59 & 0.5 &  & 0.40 & 0.8 &  & 4.0 \\
   & 1541.29145 & 37.04 & 0.5 &  & 0.75 & 1.5 &  & 3.8 \\
   & 1541.33793 & 36.46 & 0.5 &  & 0.27 & 0.5 &  & 3.8 \\
   & 1541.40453 & 36.39 & 0.5 &  & 0.20 & 0.4 &  & 3.7 \\
   & 1541.44922 & 37.01 & 0.5 &  & 0.82 & 1.6 &  & 3.5 \\
   & 1541.47928 & 37.07 & 0.5 &  & 0.86 & 1.7 &  & 3.9 \\
   & 1542.18976 & 36.75 & 0.5 &  & 0.56 & 1.1 &  & 4.7 \\
   & 1542.24088 & 35.59 & 0.5 &  & -0.60 & 1.2 &  & 4.9 \\
   & 1542.28896 & 36.85 & 0.5 &  & 0.66 & 1.3 &  & 5.6 \\
   & 1542.33138 & 36.59 & 0.5 &  & 0.40 & 0.8 &  & 4.7 \\
   & 1542.38990 & 36.35 & 0.5 &  & 0.16 & 0.3 &  & 4.6 \\
   & 1542.43474 & 36.59 & 0.5 &  & 0.40 & 0.8 &  & 4.7 \\
46 & 460.34582 & 36.97 & 0.5 & 36.19 & 0.78 & 1.6 & &3.8\\
  & 792.31009 & 36.91 & 0.5 &  & 0.72 & 1.4 & -0.06 &3.9 \\
  & 794.39999 & 36.90 & 0.5 &  & 0.73 & 1.4 & -0.07 & 4.1\\
  & 794.40812 & 36.73 & 0.5 &  & 0.54 & 1.1 & -0.24 & 4.2\\
  & 794.41615 & 36.91 & 0.5 &   & 0.72 & 1.4 & -0.06 & 4.2 \\
49$^r$ & 460.34517 & 37.22 & 3.4 & 36.02 & 1.2 & 0.4& &4.8 \\
   & 792.31824 & 34.75 & 2.0 &   & -1.27 & -0.7 &-2.47& 4.4\\
60 & 460.36269 & 37.00 & 0.8 & 36.43 & 0.57 & 0.7 & &3.1\\
   & 792.32645 & 36.70 & 0.8 &       & 0.27 & 0.3 & -0.3& 3.9\\
64 & 460.37087 & 36.23 & 0.9 & 36.33 & -0.1 & -0.1 & &3.2\\
   & 792.33456 & 36.09 & 0.7 &       & -0.24 & -0.3 & -0.14 & 3.4\\
83 & 792.35098 & 37.66 & 0.5 & 37.14 & 0.52 & 1.0 & & 4.8\\
   & 794.42461 & 37.84 & 0.5 &       & 0.70 & 1.4 & 0.18& 3.9\\
88A & 460.39414 & 39.07 & 0.5 & 36.61 & 2.46 & 4.9 & & 0.5a\\
    & 792.35970 & 39.13 & 0.5 &       & 2.52 & 5.0 & 0.06&0.5a \\
88B & 460.40029 & 35.04 & 0.5 & 36.61 & -1.57 & -3.2 & & 2.8\\
   & 792.36787 & 36.13 & 0.5 &    & -0.48 & 1.0 & 1.09& 1.4 \\
98$^r$ & 460.40542 & 39.19 & 3.6 & 36.77 & 2.42 & 0.7 & &5.0\\
   & 792.37262 & 36.91 & 1.65 &     & 0.14 & 0.1 & -2.28 & 3.7\\
101 & 460.41391 & 38.22 & 1.0 & 37.85 & 0.37 & 0.4 & &4.3 \\
    & 792.38320 & 37.74 & 0.5 &       & -0.11 & -0.2 & -0.48 & 4.4\\
115 & 460.43029 & 38.91 & 0.5 & 37.32 & 1.59 & 3.2 & &3.6\\
    & 792.40683 & 39.06 & 0.5 &       & 1.74 & 3.5 & 0.17 & 3.5\\
119 & 460.43856 & 37.59 & 0.5 & 37.14 & 0.45 & 0.9 & & 4.4\\
    & 792.41942 & 37.68 & 0.5 &       & 0.54 & 1.1 & 0.09& 4.0\\
126$^r$ & 460.44699 & 37.29 & 2.6 & 37.76 & -0.47 & -0.2 & & 6.0\\
129 & 460.45539 & 39.06 & 0.5 & 38.04 & 1.02 & 2 & &3.6\\
    & 792.42803 & 38.92 & 0.5 &       & 0.88 & 1.8 & -0.14 & 4.0\\
132 & 793.50806 & 38.83 & 0.5 & 37.81 & 1.02 & 2.0 & & 4.3\\
\hline
\end{tabular}
\end{table*}

\setcounter{table}{4}
\begin {table*}
\centering
\caption{Heliocentric velocities (contd.)}
\begin{tabular}{lcccccccc}\hline
RHy & MJD & V$_{hel}$ & $\epsilon_V$ & V$^P$ & $\delta V^P_{hel}$ & $\delta V / \epsilon_V$ & $\Delta V_{epoch}$ & H$\alpha$ EW\\
  & 50000+& kms$^{-1}$& kms$^{-1}$& kms$^{-1}$ & kms$^{-1}$& &kms$^{-1}$ & \AA \\
\hline\hline
143 & 460.46358 & 38.62 & 0.5 & 37.89 & 0.73 & 1.5 & & 3.0\\
    & 792.43758 & 38.51 & 0.5 &       & 0.62 & 1.2 & -0.11& 6.6 \\
158 & 460.47216 & 39.45 & 0.5 & 38.71 & 0.74 & 1.5 & &4.1\\
    & 792.44589 & 37.63 & 0.5 &       & -1.08 & -2.2 & -1.82 & 4.2\\
162$^r$ & 460.48031 & 38.09 & 5.0 & 37.83 & 0.26 & 0.05 & & 4.7\\   
    & 792.47709 & 37.87 & 2.5 &       & 0.04 & 0 & -0.22 & 5.3\\
163 & 792.48866 & 38.46 & 0.7 & 38.30 & 0.16 & 0.2 & & 3.2\\
165$^r$ & 793.47003 & 38.51 & 2.1 & 37.82 & 0.69 & 0.3 & & 7.7 \\
165$^r$ & 793.47003 & 38.51 & 2.1 & 37.82 & 0.69 & 0.3 & & 7.7 \\
182 & 792.49718 & 39.27 & 0.5 & 38.59 & 0.68 & 1.4 & & 4.0\\
199 & 792.50523 & 38.81 & 0.7 & 38.26 & 0.55 & 0.7 & & 4.1\\
200 & 792.51342 & 39.83 & 0.5 & 39.00 & 0.83 & 1.6 & &3.6\\
202 & 792.52180 & 38.71 & 0.5 & 38.31 & 0.4 & 0.8 & & 3.0\\
219 & 792.52990 & 39.27 & 0.5 & 38.73 & 0.54 & 1.1 & & 4.3\\
221 & 792.53800 & 40.55 & 0.5 & 38.63 & 1.92 & 3.8 & & 2.9\\
230 & 793.24775 & 39.68 & 1.4 & 38.74 & 0.94 & 0.7 & & 5.2\\
    & 794.44496 & 39.80 & 1.6 &       & 1.06 & 0.7 & 0.12& 4.2 \\
231 & 792.55084 & 39.10 & 0.5 & 38.63 & 0.47 & 0.9 & & 5.8\\
240A & 793.26704 & 39.73 & 1.3 & 38.90 & 0.83 & 0.6 & & 6.0\\
240B & 793.27862 & 38.23 & 0.5 & 38.90 & -0.67 & 1.3 & & 4.0\\
242 & 793.50000 & 40.45 & 0.5 & 39.16 & 1.26 & 2.5 & & 3.1\\
244 & 460.24334 & 37.96 & 0.5 & 38.54 & -0.58 & -1.2 & &4.7\\
260 & 460.26471 & 40.59 & 0.5 & 39.28 & 1.31 & 2.6 & & 3.6\\
    & 793.28678 & 40.58 & 0.5 &      & 1.30 & 2.6 & -0.01& 4.3 \\
281 & 793.29503 & 33.32 & 1.3 & 39.22 & -5.80 & -4.5 & & 4.6\\ 
    & 794.45771 & 32.80 & 1.1 &     & -6.42 & -5.8 & -0.52& 22.8 \\
    & 1541.18724 & 32.66 & 1.2 &   & -6.56 & -5.5 & -0.66& 6.7 \\
    & 1541.34444 & 33.19 & 0.6 &   & -6.03 & -10.1& -0.13& 6.1 \\
    & 1542.33699 & 32.77 & 1.2 &   & -6.45 & -5.4 & -0.55& 5.5 \\
297 & 460.27310 & 40.29 & 0.5 & 39.66 & 0.63 & 1.3 & & 4.1\\
    & 793.30331 & 40.33 & 0.5 &       & 0.67 & 1.3 &0.04& 4.7\\
    & 794.49131 & 40.41 & 0.5 &       & 0.75 & 1.5 & 0.12& 4.3\\ 
298 & 460.28963 & 40.63 & 0.5 & 39.68 & 1.05 & 2.1 & & 3.3\\
    & 793.31490 & 40.61 & 0.5 &       & 1.03 & 2.1 & -0.02& 4.1 \\
301 & 460.48851 & 40.98 & 1.0 & 39.77 & 1.2 & 1.2 & & 5.0\\
    & 793.32307 & 40.08 & 0.5 &       & 0.31 & 0.6 & -0.9 & 6.1 \\
309 & 460.49697 & 40.82 & 0.5 & 39.15 & 1.67 & 3.1 & & 4.9\\
331 & 793.33144 & 40.13 & 0.5 & 39.28 & 0.85 & 1.7 & & 3.0\\
346 & 793.34021 & 40.90 & 0.5 & 40.16 & 0.74 & 1.5 & & 3.5\\ 
367 & 793.34828 & 41.10 & 0.5 & 40.24 & 0.86 & 1.7 & & 4.8\\
369 & 793.35632 & 41.51 & 0.5 & 40.30 & 1.21 & 2.4 & & 2.7\\
371 & 793.36490 & 38.23 & 0.7 & 39.82 & -1.59 & -2.3 & & 3.8\\
376 & 793.51613 & 41.00 & 0.5 & 39.99 & 1.01 & 2.0 & & 3.8\\
377 & 793.37344 & 42.40 & 0.5 & 40.38 & 2.02 & 4.0 & & 3.8\\
386 & 793.40658 & 41.30 & 0.5 & 40.08 & 1.22 & 2.4 & & 4.4\\
    & 794.50953 & 41.58 & 0.8 &       & 1.50 & 1.9 & 0.28& 4.8\\
390 & 793.48502 & 40.94 & 1.6 & 39.84 & 1.10 & 0.7 & & 8.7\\
391 & 793.42219 & 39.69 & 0.8 & 39.61 & 0.08 & 0.1 & & 5.1 \\
399 & 793.43026 & 40.76 & 0.5 & 39.90 & 0.86 & 1.7 & &3.8 \\
402 & 793.43834 & 40.92 & 0.5 & 39.92 & 1.0 & 2.0 & & 2.9\\
    & 794.53181 & 39.97 & 0.8 &       & 0.05 & 0 & -0.95 & 3.8\\
\hline
\end{tabular}
\end{table*}

\setcounter{table}{4}
\begin {table*}
\centering
\caption{Heliocentric velocities (contd.)}
\begin{tabular}{lcccccccc}\hline
RHy & MJD & V$_{hel}$ & $\epsilon_V$ & V$^P$ & $\delta V^P_{hel}$ & $\delta V / \epsilon_V$ & $\Delta V_{epoch}$ & H$\alpha$ EW\\
  & 50000+& kms$^{-1}$& kms$^{-1}$& kms$^{-1}$ & kms$^{-1}$& &kms$^{-1}$ & \AA \\
403 & 793.45354 & 42.96 & 0.5 & 40.11 & 1.75 & 3.5 & & 5.2\\
    & 794.54738 & 73.76 & 0.8 &       & 33.7 & 42 & 30.80& 5.7\\
    & 1541.19393& 69.42 & 0.6 &      & 29.3 & 49 & 26.46& 7.8 \\
    & 1541.24538& 75.41 & 0.5 &      & 35.3 & 71 & 32.45 &7.5 \\
    & 1541.29500& 78.83 & 0.5 &      & 38.7 & 77 & 35.87 &6.2 \\
    & 1541.35213& 79.99 & 0.8 &      & 39.9 & 50 & 37.03 & 8.1 \\
    & 1541.40799& 78.46 & 0.5 &      & 38.4 & 77 & 35.50 & 6.4\\
    & 1541.45265& 75.15 & 0.6 &      & 35.0 & 58 & 32.19 & 5.9\\
    & 1541.48272& 72.16 & 0.5 &      & 32.1 & 64 & 29.20 & 3.9\\
    & 1542.19549& 20.23 & 0.5 &      & -19.8& -40& -22.69 & 7.0 \\
    & 1542.24438& 27.69 & 0.5 &      & -12.4& -25&-15.27 & 5.9 \\
    & 1542.29250& 37.20 & 0.5 &      & -2.91& -6 &-5.76 & 7.2 \\
    & 1542.34197& 47.04 & 0.6 &      & 6.9 & 12 & 4.08 & 6.4 \\
    & 1542.39357& 56.73 & 0.5 &      & 16.6& 33& 13.77 &6.2\\
    & 1542.43815& 64.10 & 0.5 &        & 24.0& 48& 21.14&6.2 \\
G8-17 & 793.39242 & 36.85 & 0.5 & 35.96 & 0.91& 1.8&  & 9.8\\ 
\hline
\end{tabular}
\end{table*}

\section {Discussion}

\begin{figure}
\psfig{figure=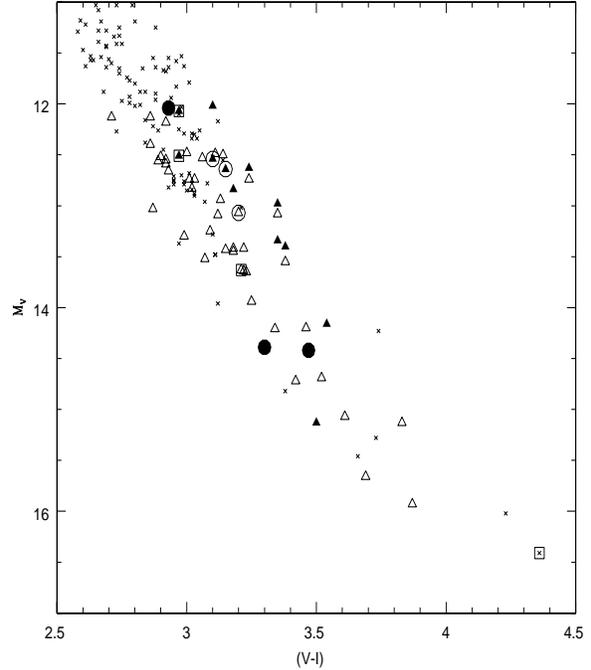,height=9cm,width=8cm
,bbllx=8mm,bblly=57mm,bburx=205mm,bbury=245mm,rheight=10cm}
\caption{The (M$_V$, (V-I)) colour-magnitude diagram for late-type Hyades dwarfs. Crosses are stars
lacking HST observations; open triangles are stars unresolved by HST (RG97); solid triangles
are resolved HST binaries (RG97); solid points mark the locations of RHy 42, 281 and 403;
the three possible velocity variables, RHy 158, 221 and 377, are identified using open circles,
while open squares identify four stars with unusual cross-correlation profiles: RHy 49, 244, 162 and 165.}
\end{figure}

\subsection {New spectroscopic binaries}

\begin{figure}
\psfig{figure=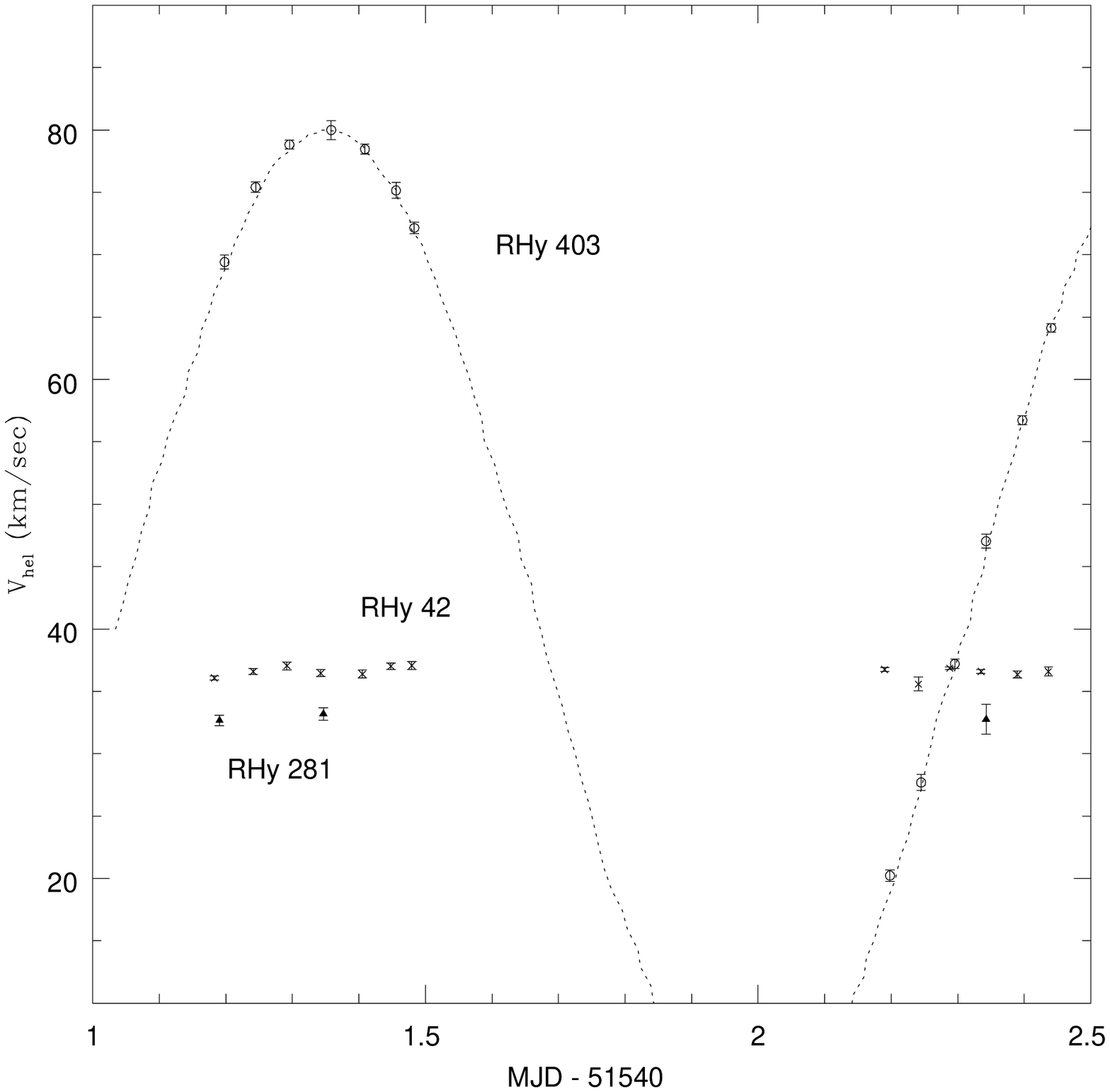,height=9cm,width=8cm
,bbllx=8mm,bblly=57mm,bburx=205mm,bbury=245mm,rheight=10cm}
\caption{Velocity measurements from December, 1999 for RHy 42 (crosses), RHy 281 (triangles) 
and RHy 403 (open circles). The curve matched to the last set of observations is a 
sine-wave with P=1.276 days, amplitude $\pm 40$ kms$^{-1}$ and T$_0$=51541.3521 days (MJD).}
\end{figure}

Figure 2 plots the colour-magnitude diagram for lower main-sequence in the Hyades,
marking the spectroscopic binaries and candidate binaries identified 
from our observations. Three stars in Table 4 are prime binary candidates: RHy 42, 
281 and
403. Two are confirmed as binaries based on the 1997 data alone: RHy 42,
a double-lined system; and RHy 403, a single-lined system, whose 
radial velocity increased by 31 kms$^{-1}$ in only 26 hours in December, 1997. 
The third star, RHy 281,  exhibits no variation in velocity, but the 
measured value is offset by $\sim5 \sigma$ from the expected
value for a cluster member. These three stars were therefore chosen as the focus 
of our most recent observations,

Figure 3 plots the results of our December, 1999 observations.
RHy 403 is confirmed as a spectroscopic binary. The curve plotted is a simple
sinewave with a period of 1.276 days and an amplitude of 40 kms$^{-1}$, matched by
eye to the data.
Phase zero is at MJD=51541.3521 days. The uncertainty in the derived period is $\pm 0.005$ days.
The December 1997 observations are consistent with being obtained at phases
of $\phi = 0.487$ and $\phi = 0.339$ respectively, with $\phi = 0.25$ at 40794.429.
The total number of cycles between the latter date and the observed maximum on
December 29 can only be given as 584$\pm 2$.

\begin{figure*}
\psfig{figure=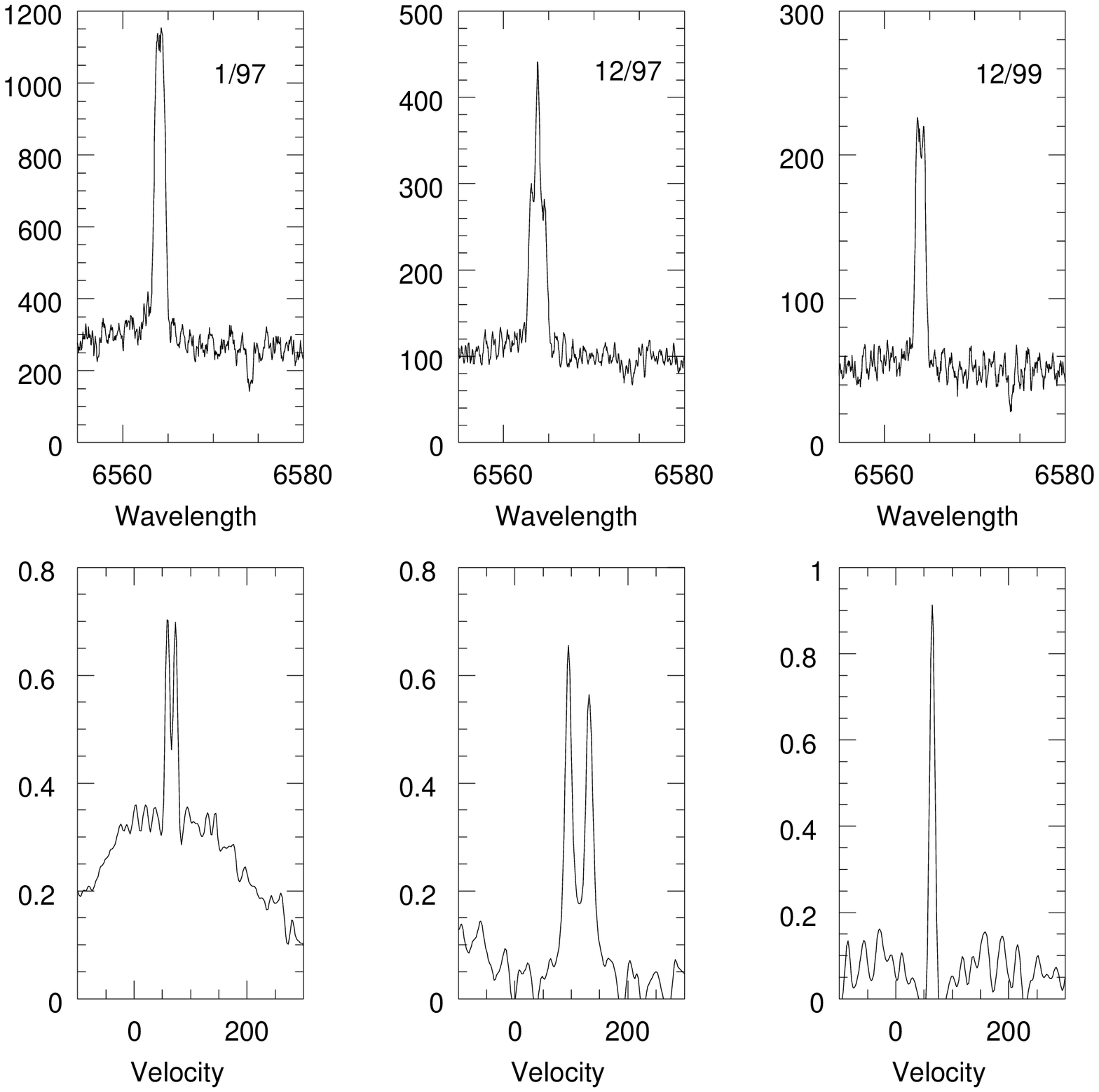,height=9cm,width=15cm
,bbllx=8mm,bblly=57mm,bburx=205mm,bbury=245mm,rheight=10cm}
\caption{H$\alpha$ emission and typical cross-correlation results for 
RHy 42 observations. The left panels plot data from January, 1997; the
middle panel, the December, 1997 observation; the right
panel, data from December, 1999. The 1997 data clearly identify RHy 42 as a
double-lined binary with components of nearly equal luminosity.}
\end{figure*}

The cross-correlation functions from our 1997 observations of RHy 42 clearly
identify the star as a double-lined spectroscopic binary, as illustrated in figure 4. We
also plot H$\alpha$ emission line data: the unusual profile in December, 1997 is the
result of the superposition of two lines of almost equal strength, each
possessing a central reversal. The Ca I line at 6598\AA\ is also clearly
double at the latter epoch. However, there is no evidence for either velocity
variations or line-splitting in the December, 1999 data.

These observations are consistent with the identification of RHy 42 as a 
moderate-period binary system. The semi-major axis in an equal-mass
binary is given by
\begin{displaymath}
a = {{{\mathcal{G}} (M_1 + M_2)} \over v_{circ}^2}
\end{displaymath}
Both components have M$_V \sim 12.5$, or, from Henry \& McCarthy's (1993) (mass-M$_V$) relation,
 $M \sim 0.2 M_\odot$. If we assume that the observed velocity difference in December, 1997, 
$\delta V = 34$ kms$^{-1}$, is close to the maximum value, and an inclination of 45$^o$, then
$v_{circ} \sim 24$kms$^{-1}$ and $P \sim 0.75$ years. Presumably our 1999 observations
happened to catch the system at phase $\phi = 0^o$ or $180^o$, with both
stars having radial velocities close to the systemic value.

Although RHy 281 has a radial velocity offset by 6 kms$^{-1}$ from
that expected for a Hyades star, our data show no evidence for significant velocity variation. 
The rms dispersion of all five observations, spanning two years, is only 0.29 kms$^{-1}$. 
The average H$\alpha$ emission
properties of this star are entirely consistent with Hyades membership. It is
possible that RHy 281 may be no longer be bound to the Hyades cluster
proper, although still part of the more extensive moving group. Alternatively,
RHy 281 may be a long-period binary system, with an
as-yet undetected low-mass companion. Further observations of this star are clearly warranted.

Seven other stars are identified as candidate binaries: three on the basis of the velocity
data listed in table 4; four display anomalies in the shape of the
cross-correlation peak, with one star almost certainly binary.
In the former group, 
RHy 221 and 377 both have only a single velocity determination, but
in both cases, $\delta V / \epsilon_V \sim 4$. Both are  known
as a binary from HST observations. The peak in RHy 377's cross-correlation function is 
slightly asymmetric towards lower velocities, suggesting that the known
companion may be responsible for the offset from the predicted velocity.
The third star is RHy 158.
Our two velocity determinations differ by only 1.8 kms$^{-1}$
($\sim 4 \epsilon_V$), but
this star was also observed by Terndrup et al (2000). Their datum 
differs significantly from both our measurements.

RHy 49, 162, 165 and 244 all have cross-correlation functions with maxima
suggestive of double-lined systems, with velocity separations of 5 to 10 kms$^{-1}$
(figure 5). The first three stars are all rapid rotators, and it is possible that 
rapid rotation, rather than binarity, may be responsible for the
anomalous structure (although RHy 49 is resolved as binary by HST). 
RHy 244, however, has low $v \sin(i)$, and the
narrow cross-correlation peak has blueward asymmetry, suggesting a
contribution from a lower-luminosity companion (see, for comparison, 
figures 3 and 7 in Stauffer et al, 1997). We know from HST imaging that
RHy 244 has a companion at $\Delta \sim 0.1$ arcseconds, $\delta I \sim 1.1$ 
magnitudes. The observed asymmetry is not inconsistent with the expected
velocity difference of these known components. 

\begin{figure}
\psfig{figure=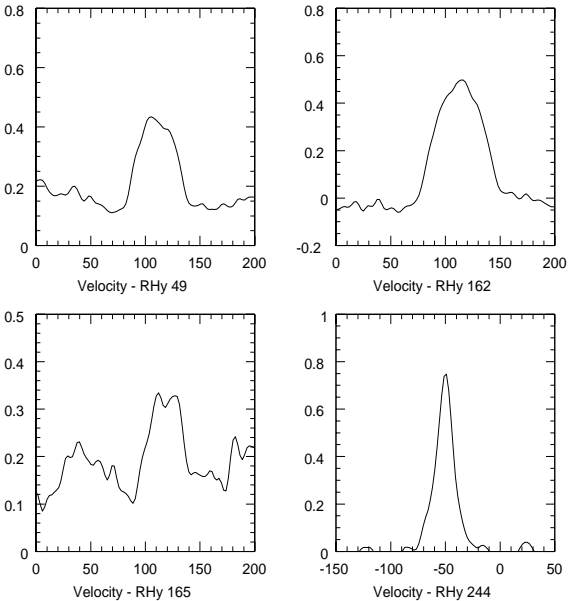,height=8cm,width=8cm
,bbllx=8mm,bblly=57mm,bburx=205mm,bbury=245mm,rheight=9cm}
\caption{The central regions of the cross-correlation functions for RHy 49,
162, 165 and 244. The RHy 244 cross-correlation peak has an asymmetric
blue wingm, while the remaining three functions exhibit anomalous profiles. }
\end{figure}

\subsection {The companion of RHy 403} 

Our observations of RHy 403 provide sufficient information to
set firm constraints on the mass of the companion. If we assume a circular orbit, as 
suggested by the sine curve matched against the data in figure 3, then the
observed maximum velocity (80 kms$^{-1}$) and V$^P \sim 40$kms$^{-1}$ imply
an orbital velocity, $v sin(i) = 40$kms$^{-1}$, and a projected
semi-major axis, $a sin(i) \sim 701,000$ km, or 0.0047 A.U. The systemic
mass function is given by
\begin{displaymath}
f(m) \ = \ {{M^3_2 sin^3(i)} \over {(M_1 + M_2)^2}} \ = \ {{(a sin(i))^3} \over P^2}
\end{displaymath}
where $P$ is the period, 1.275 days.  Substituting for $P$, we derive $f(m) = 0.0085$.

The fact that RHy 403 is a single-lined system allows us to estimate both a lower limit
and an upper limit to M$_2$, the mass of the unseen companion. First, we can assume
that the primary contributes 
most of the flux, even at 8000\AA. Hence, the observed absolute magnitude provides
an estimate of the mass of the primary, RHy 403A. Based on (BC$_I$, (V-I)) data from
Leggett et al (1996) and the photometry given in Table 1, we derive 
M$_{bol} \sim 11$. Matched against the 600 Myr-old models computed by Burrows 
et al (1993), this implies a mass of $\sim0.15 M_\odot$. 
Substituting for $M_1$ in the mass function gives
\begin{displaymath}
M_2^3 sin^3(i) \ = \ 1.9 \times 10^{-4} + 0.0085 M_2^2
\end{displaymath}
Setting $i = 90^o$ gives a lower limit to the mass of the companion, $M_2 > 0.062 M_\odot$,
the first solution of the cubic equation.

In principle, RHy 403 could have an evolved companion; in practice, we
can effectively rule out this option.  With an age of 625 Myrs, the 
lowest luminosity white dwarfs in the Hyades have M$_V \sim 12$ and T$_{eff} \sim 15,000$K.
Unresolved Hyades M dwarf/white dwarf binaries therefore have unusual 
colours, as demonstrated by HZ 9 (M2 + DA: M$_V = 10.6,$ (V-I)=1.8). 
A neutron star companion, $M_2 = 1.4$ to 2 M$_\odot$, requires $10^o.5 > i > 9.3$ and
$0.026 < a < 0.029$ AU; the extensive mass loss during a supernova excludes the possibility of
a primordial binary at these separations, while a system forming post-SN would still
be expected to be bright at X-ray wavelengths. Stern et al (1994) do not detect a ROSAT
source close to RHy 403 in their survey of the cluster.

Given that the companion is either a low-luminosity main-sequence star or a brown dwarf, we
can use the non-detection in both the direct spectrum and, more stringently, the cross-correlation
analysis, to set an upper limit to M$_2$. A 0.06 M$_\odot$ brown dwarf is predicted to
have a temperature of $\sim 1900$K at age 625 Myrs, corresponding to a spectral type of L2
and a luminosity of $10^{-4} L_\odot$.
With an absolute magitude of M$_I \sim 15$, 
over 4 magnitudes fainter than RHy 403A, such an object lies well below our detection limit.

\begin{figure}
\psfig{figure=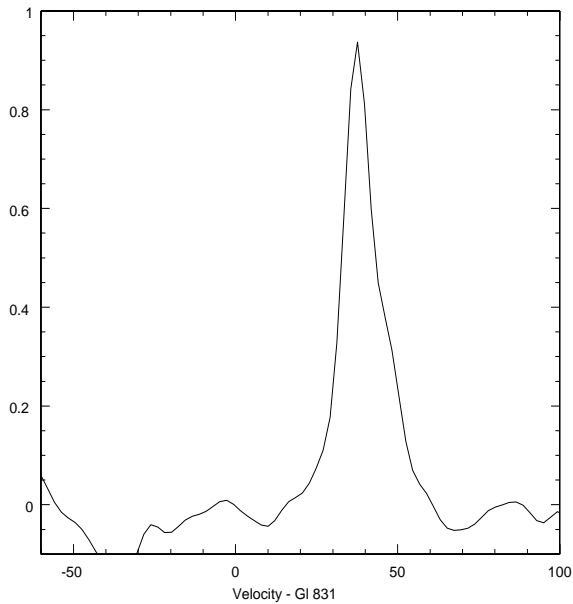,height=8cm,width=8cm
,bbllx=8mm,bblly=57mm,bburx=205mm,bbury=245mm,rheight=9cm}
\caption{The cross-correlation peak for Gl 831AB matched against our template, Gl 447. The peak
is asymmetric towards higher velocities. We interpret the asymmetry as due to
the known companion, Gl 831B.}
\end{figure}

How faint a companion could we expect to detect? We can address this question in two ways.
First, we have modelled
an unequal-mass, late-type M-dwarf binary by combining  appropriate spectra spanning a
range of flux ratio. Based on those tests, we would expect to detect anomalies 
in the cross-correlation function for flux ratios exceeding 4:1 at $\lambda > 8000\AA$.
Second, we can use observations of known spectroscopic binaries to infer detection limits.
The data  plotted in figure 5 suggest that we have detected RHy244B, $\delta I = 1.1$ magnitude, 
a flux ratio of $\sim 3:1$. Similarly,  
Gl 831 is a known binary, with M$_V(A) = 12.5$, M$_V(B) = 15.5$;
masses, $M_A = 0.25 M_\odot, M_B = 0.1 M_\odot$; $a \sim 0.21$ arcseconds; and P$\sim 1.92$ years
(Henry, 1990). Cross-correlating our spectra of Gl 831 against data for Gl 447 reveals
an asymmetric peak (figure 6). Given $\Delta$M$_V \sim 3$ magnitudes and
$\Delta$M$_K \sim 1.3$ magnitudes for the two components, the likely magnitude
difference at I is close to 2 magnitudes, or a flux ratio of $\sim$6:1. 
Our data for RHy 403 have lower signal to noise, but it is unlikely that
a companion with a flux ratio exceeding 4:1 would have escaped detection.

Our flux limit corresponds to an absolute magnitude of M$_I = 12.3$.  
Matched against the Burrows et al (1993) models, this implies an upper 
mass limit of $M_2 <  0.095 M_\odot$ for RHy 403B. 

The lowest-mass member of the Hyades currently known is LH0418+13 (Reid \& Hawley, 1999), 
with M$_I \sim 14$ and a mass of approximately 0.08 M$_\odot$. There is a probability of
72\% that the orbit of RHy 403B has $i > 50^o$ and $M < 0.08 M_\odot$; the probability is
66\% that $i > 55^o$ and $M < 0.075 M_\odot$. Thus, RHy 403B is the first strong 
candidate for a brown dwarf member of the Hyades cluster. Unfortunately, since $a sin(i) = 0.0047$AU
and $r = 53$ parsecs, the maximum expected angular separation (for ${M_2 \over M_1} \sim 2.5$)
is only 0.25 mas. High-resolution spectroscopy at near-infrared wavelengths may provide stronger 
constraints on the upper mass limit.

\subsection{The binary fraction of late-type Hyades M dwarfs}

Prior to our observations, ten of the fifty-one Hyades stars listed in
Table 1 were known to be in binary systems. Our spectroscopic data add three definite
new systems, RHy 42, 158 and 403, with RHy 281 also a strong candidate binary.
In addition, RHy 162 and 165 show indications of duplicity, although the
high intrinsic rotation of these stars render less certain the interpretation of the cross-correlation
data. RHy 244, 377 and, perhaps, 49 are also possible spectroscopic binaries, but all
three were identified previously as binaries based on the HST imaging. Our present
radial velocity data are insufficient to determine whether the spatially-resolved
and hypothetical spectrscopic components are one and the same. 

Forty-eight of the fifty-one M dwarfs listed in Table 1 have absolute magnitudes
M$_V > 12$. Amongst the 47 systems (RHy 240A and B are observed separately)
at least 11, and perhaps as many as 14,
are binary or multiple systems - a multiplicity fraction of 23 to 30\%, with
a formal uncertainty of 7\%. In comparison, the binary fraction amongst 
the (more completely surveyed) M dwarfs
in the 8-parsec sample is 33\% (Reid \& Gizis, 1997a). 

Further analysis is limited by the small numbers in  both the Hyades and field
M dwarf samples. However, Table 6 compares the distribution of projected 
separations, $\Delta$, adopting $\Delta = 0.8 a$ for nearby systems with known
orbits. We list statistics for all 8-parsec binaries with M-dwarf primaries and for the subset
where the primary has $M < 0.3 M_\odot$. Within the admittedly large uncertainties, the
field star and cluster distributions are identical.

\begin {table*}
\centering
\caption{Distribution of separation of binary components}
\begin{tabular}{lcccccc}\hline 
 & Hyades & & Field & all M & Field & $M < 0.3 M_\odot$ \\
\hline\hline
$\Delta$ & N & \% & N & \% & N & \% \\
\hline
$\le 10$ AU & 7 & 50\% & 17 & 50\% & 11 & 65\% \\
10 - 100 AU & 5 & 36 & 11 & 32 & 6 & 35 \\
$>100$ AU & 1 & 14 & 6 & 18 & 0 & 0 \\
\hline
\end{tabular}
\end{table*}

\section{Chromospheric activity in low-mass Hyades dwarfs}

\begin{figure}
\psfig{figure=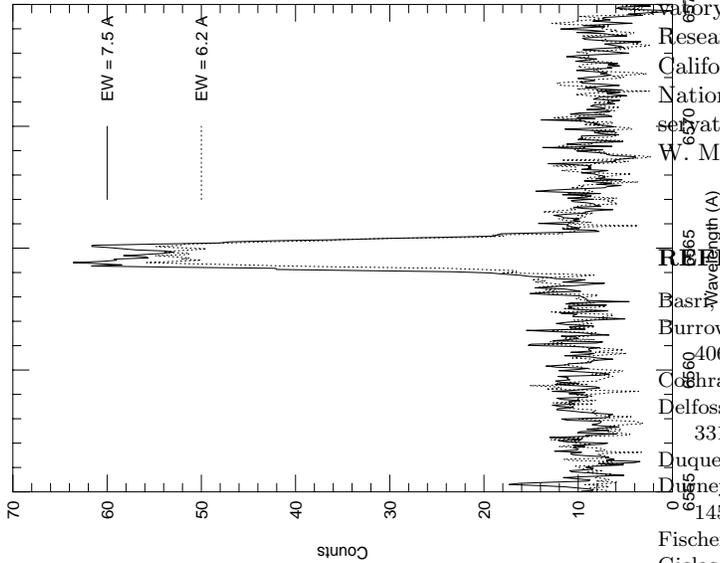,height=8cm,width=8cm
,bbllx=8mm,bblly=57mm,bburx=205mm,bbury=245mm,rheight=10cm}
\caption{H$\alpha$ line profiles from successive observations of RHy 403 (MJD 1541.245 and 1541.295).}
\end{figure}

\begin {table*}
\centering
\caption{Rotation and activity in low-mass Hyades stars}
\begin{tabular}{lccclccc}\hline 
 Name & v sin(i) & M$_{bol}$ & Log (L$_\alpha$/L$_{bol}$) &  
Name & v sin(i) & M$_{bol}$ & Log (L$_\alpha$/L$_{bol}$)   \\
\hline\hline
RHy 9 & 3.5 & 10.55 & -3.88  & RHy 23 & 4 & 10.08 & -3.86 \\
42 & 3.5 & 10.19 & -3.99 & 46 & 5 & 10.28 & -3.89 \\
49 & 22 & 9.81 & -3.80 & 60 & 19 & 9.66 & -3.95 \\
64 & 19 &  10.68 & -4.01 & 83 & 8 & 11.29 & -3.98 \\
98 & 23 & 10.68 & -3.87 & 101 & 16 & 10.89 & -4.14 \\
115 & 7 & 10.05 & -3.87 & 119 & 7.5 & 10.73 & -3.98 \\
126 & 21 & 10.36 & -3.82 & 129 & 3.5 & 9.94 & -3.85 \\
143 & 4 & 10.15 & -3.85 & 158 & 11 & 10.14 & -3.92 \\
162 & 35 & 10.59 & -3.77 & 163 & 19 & 9.83 & -3.94 \\
165 & 17 & 11.67 & -3.99 & 182 & $<3$ & 9.80 & -3.87 \\
199 & 16 & 10.46 & -3.99 & 200 & 7.5 & 10.06 & -3.84 \\
202 & 2.5 & 10.53 & -4.04 & 219 & 7 & 10.55 & -3.76 \\
221 & 2.5 & 10.50 & -3.98 & 230 & 16 & 11.47 & -3.91 \\
231 & 6.5 & 9.69 & -3.73 & 240A & 14.5 & 11.68 & -3.90 \\
240B & 14.5 & 11.72 & -4.09 & 242 & 6 & 10.01 & -3.93 \\
244 & 2.5 & 9.46 & -3.76 & 260 & 2.5 & 10.77 & -3.90 \\
281 & 9.5 & 11.32 & -3.66 & 297 & 13 & 10.52 & -3.80 \\
298 & 6 & 10.10 & -3.86 & 301 & 12 & 10.57 & -3.72 \\
331 & 6.5 & 10.66 & -3.94 & 346 & 3.5 & 10.44 & -3.98 \\
367 & 3 & 10.11 & -3.82 & 369 & $< 2.5 $ & 10.05 & -3.99 \\
371 & 14 & 9.23 & -3.90 & 376 & 4 & 10.00 & -3.83 \\
377 & 9 & 9.79 & -3.90 & 386 & 9 & 12.00 & -3.95 \\
390 & 18 & 11.25 & -3.77 & 391 & 24 & 9.94 & -3.85 \\
399 & $< 2.5$ & 9.64 & -3.89 & 402 & 11.5 & 10.94 & -3.75 \\
403 & 5.5 & 11.10 & -3.82 \\
\hline
Jones et al \\
Br 262 & 37 & 11.40 & -4.08 & Br 804 & 31 & 11.81 & -3.86 \\
\hline
Stauffer et al \\
RHy 26 & $<6$ & 8.96 & -3.76 & RHy 28 & 11.5 & 9.45 & -3.87 \\
62 & $<6$ & 8.28 & -4.03 & 99 & $<6$ & 9.03 & -3.89 \\
100 &$<6$ & 8.93 & -4.35 & 123 & $<6$ & 8.69 & -5.09 \\
142 & 11.5 & 8.42 & -3.83 & 161 & $<6$ & 9.35 & -3.88 \\
189 & $<6$ & 9.38 & -4.66 & 203 & $<6$ & 9.09 & -5.05 \\
211 & 6 & 8.62 & -4.39 & 269 & 26 & 8.79 & -3.75 \\
283 & $<6$ & 8.53 & -3.93 & 289 & 6 & 9.07 & -4.71 \\
294 & 11 & 8.79 & -3.83 & 296 & $<6$ & 9.33 & -3.87 \\
322 & $<6$ & 9.52 & -3.35 & 349 & $<6$ & 9.11 & -3.83 \\
350 & $<6$ & 10.16 & -3.40 & 384 & 25 & 8.52 & -3.70 \\
GH 7-33 & 12.5 & 8.48 & -3.74 & Lei 44 & 18 & 8.61 & -3.63 \\
Lei 73 & $<6$ & 8.72 & -4.86 & LP 357-160 & $<6$ & 8.73 & -4.67 \\
LP 358-716 & $<6$ & 8.93 & -4.05 & LP 359-2423 & 9.5 & 8.79 & -3.76 \\
LP 359-262 & $<6$ & 8.84 & -4.55 & LP 413-93 & $<6$ & 9.13 & -4.14 \\
LP 535-101 & $<6$ & 9.09 & -4.10 & vA 54 & $<6$ & 8.94 & -3.94 \\
vA 106 & 10 & 9.16 & -3.82 \\
\hline
Terndrup et al \\
RHy 128 & 19 & 10.88 & -3.86 & RHy 138 & 22 & 11.66 & -3.96 \\
 206 & 24 & 10.47 & -3.93 & LP 416-573 & 6 & 10.18 & -4.01 \\
LP 358-735 & 25.5 & 10.16 & -3.82 & LP 476-648 & 6 & 10.18 & -3.94 \\
LP 358-717 & 18 & 10.64 & -3.86 \\ 
\hline
\end{tabular}
\end{table*}

\subsection {Rotation and activity}

Table 8 lists the rotational velocities derived for the  Hyades dwarfs in our sample. 
All are chromospherically active, with substantial H$\alpha$ emission. As indicated by
the individual measurements listed in Table 5, the level of activity can vary by 
up to 30\% on relatively short timescales. 
This intrinsic variability is the dominant source of uncertainty
in determining stellar activity from one or two observations, as is the case here.
Figure 7 illustrates this effect, comparing successive 
observations of RHy 403 from Dec 29, 1999. 
Note the central reversal in the stronger H$\alpha$
line profile. As Stauffer et al point out, the majority of slow rotators ($v \sin(i) < 12$ kms$^{-1}$)
have similar profiles, while stars with faster rotation have either flat-topped or peaked H$\alpha$ lines.

\begin{figure*}
\psfig{figure=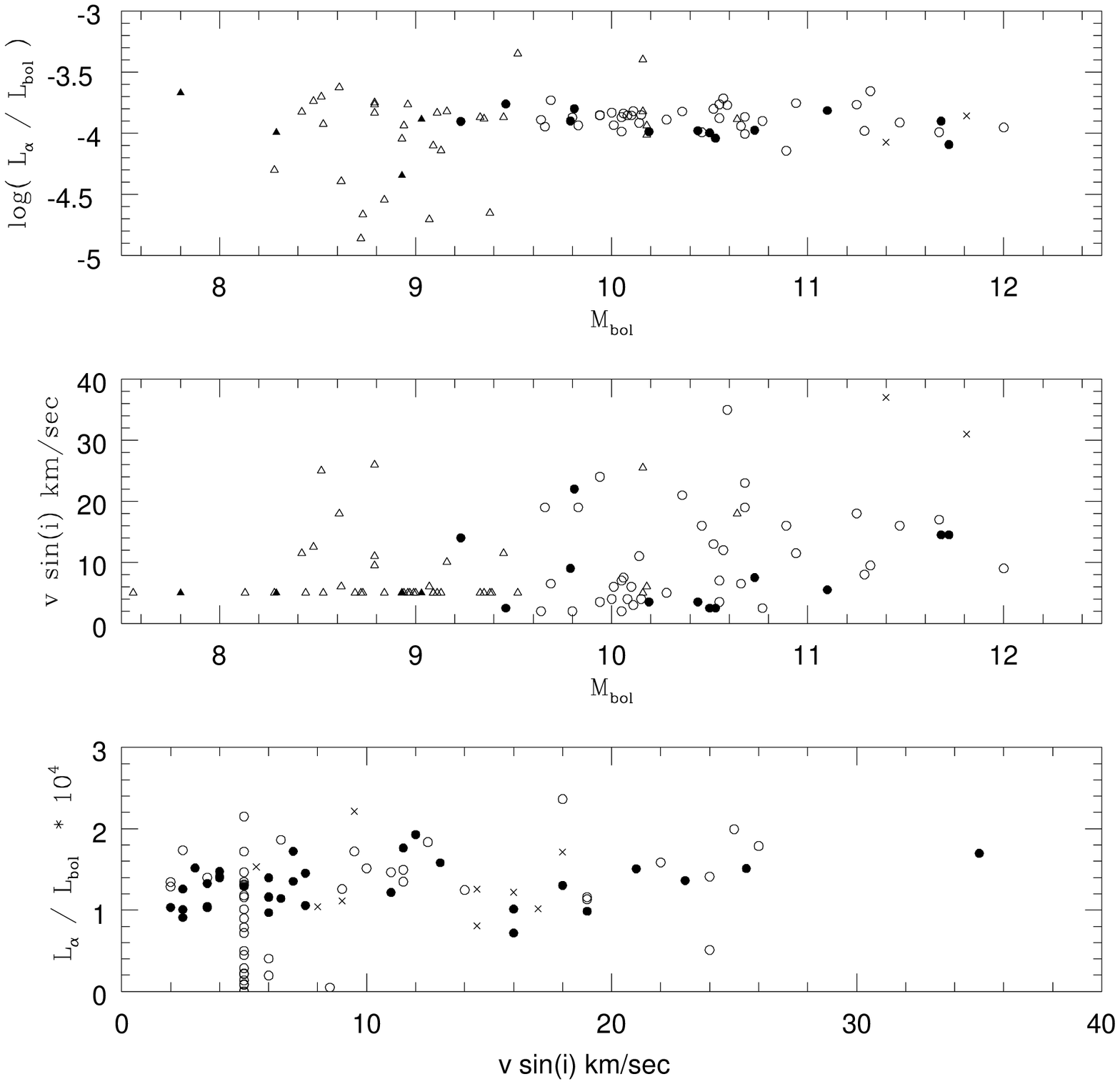,height=12cm,width=10cm
,bbllx=8mm,bblly=67mm,bburx=145mm,bbury=245mm,rheight=13cm}
\caption{ Chromospheric activity and rotation in Hyades dwarfs. The upper panel (a) plots
activity, log$L_\alpha/L_{bol}$, for M dwarfs from this paper (circles), and 
from previous studies (Stauffer et al (1997), 
Jones et al (1996), Terndrup et al (2000) - all triangles) Solid points mark known binaries.
The middle planel (b) plots the (M$_{bol}$, $v \sin(i)$ relation for the same stars, using the same
symbols. 
The lowest panel (c) plots rotation against activity, where all of the
Hyades observations are divided into three subsets based on M$_{bol}$. Solid points
mark stars with M$_{bol} < 10$; open circles have $10 < M_{bol} < 11$; the faintest stars are
marked as crosses.}
\end{figure*}

\begin{figure*}
\psfig{figure=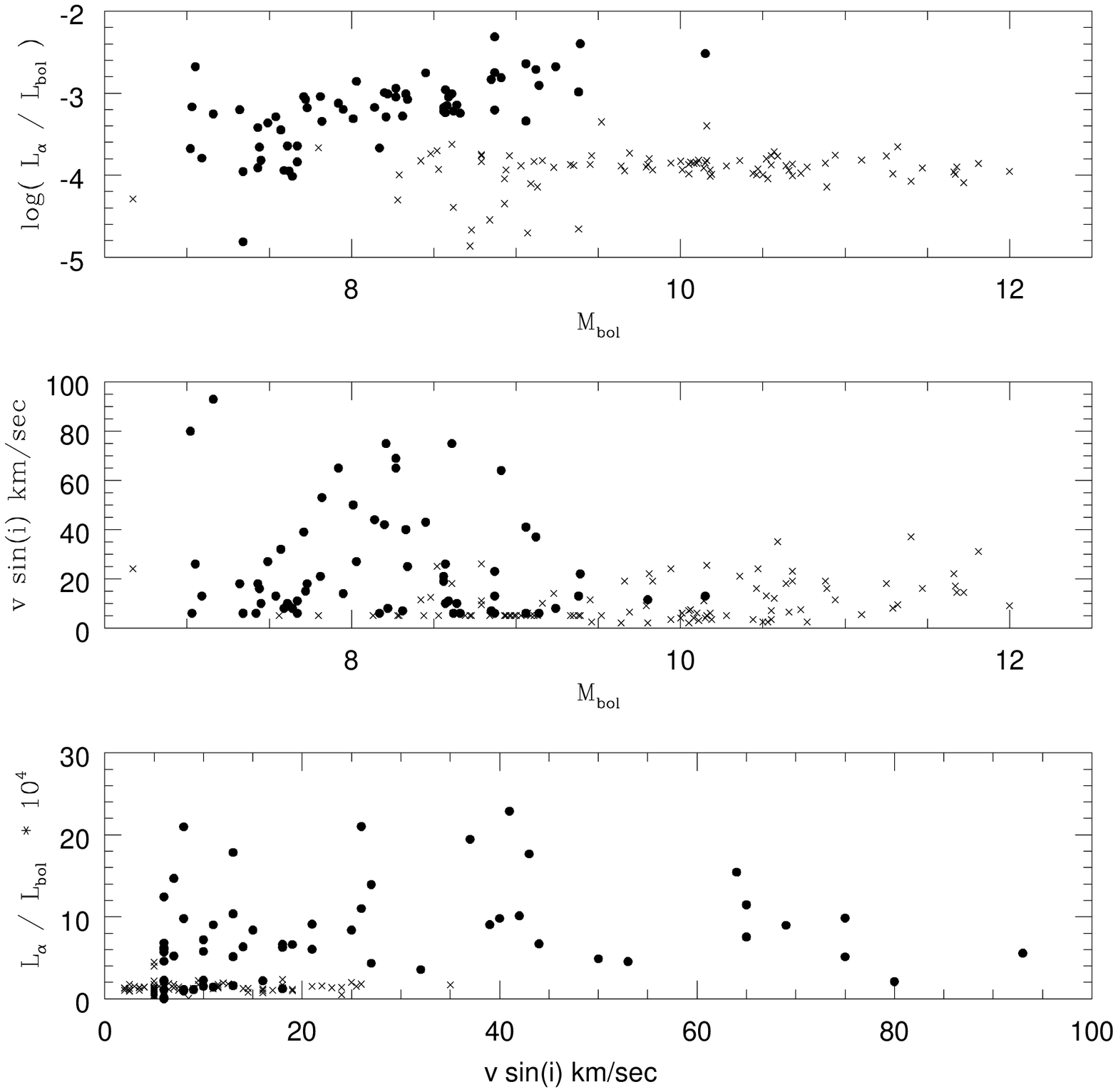,height=12cm,width=10cm
,bbllx=8mm,bblly=67mm,bburx=145mm,bbury=245mm,rheight=13cm}
\caption{H$\alpha$ emission and typical cross-correlation results for 
Activity/rotation correlations for Pleiades stars. The data are taken from Terndrup et al (2000),
Basri \& Marcy (1995) and Jones et al (1996). 
Pleiades stars are plotted as solid points; all Hyades stars
from figure 8 are plotted as crosses for comparison. }
\end{figure*}

We are interested in determing the extend of any correlation between the level of chromospheric
activity and the underlying stellar rotation. The standard paradigm (Stauffer et al, 1997)
envisions a direct correlation, with activity increasing
with increasing rotation until a ``saturation'' velocity is achieved, where the
chromosphere and coronal loops are fully active. Recent observational results, however, have
raised serious questions as to whether this scenario is maintained in low-mass, late-type 
M dwarfs (Hawley et al, 1999). 

\begin{figure*}
\psfig{figure=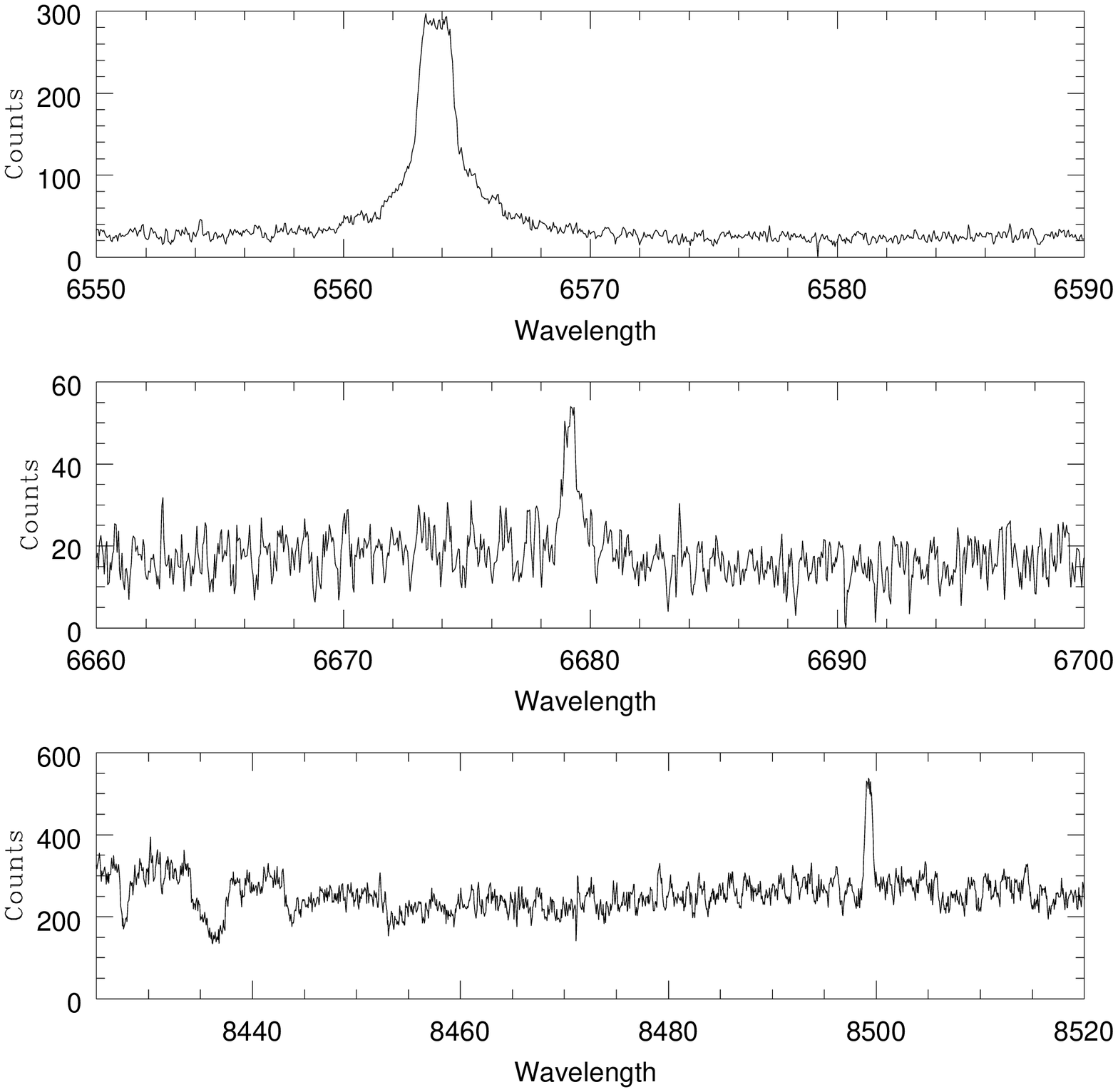,height=12cm,width=10cm
,bbllx=8mm,bblly=67mm,bburx=145mm,bbury=245mm,rheight=13cm}
\caption{H$\alpha$, He I 6678\AA\ and Ca II 8500\AA\
emission during the flare outburst of RHy 281 }
\end{figure*}

The standard rotation/activity hypothesis rests on an extension of the observed
behaviour of G and K dwarfs, where there is a clear activity-rotation correlation. 
Both parameters are believed to be age dependent, with the star spinning down as it loses angular 
momentum due to coupling of the star's magnetic field with the local medium (Skumanich, 1972).
Since activity in these stars is powered by a shell dynamo (Parker, 1955), located at the boundary
between the radiative core and the outer convective layers, the activity-rotation
correlation is expected. 
This mechanism, however, cannot be present in fully convective low-mass stars.
Indeed, early studies predicted a break in the level of activity at a mass of $\sim0.3 M_\odot$, 
where the radiative core disappears. The continued presence of high levels of
activity in these stars is probably due to the presence of a turbulent (shear) dynamo
mechanism (Durney et al, 1993). In that case, there is no expectation of a direct
correlation with rotational velocity.

Investigating this question requires a reliable measurement of the level of
activity in stars with temperatures between 4000 and 2400K. The equivalent width of the
H$\alpha$ emission line is frequently used for this purpose. This parameter, however, provides
a misleading measure of activity, since the equivalent width produced by a given line
flux depends on the level of the underlying continuum.  This is a particularly
important consideration for M dwarfs, since the R$_C$-band flux decreases rapidly with decreasing
temperature from spectral type M0 ($\sim 4000$K) through M5 ($\sim 2800$K)  to M9 ($\sim 2100$K).
This complication can be avoided by calculating the parameter $L_\alpha / L_{bol}$, the
fraction of the total luminosity emitted in H$\alpha$. This provides a measure of stellar activity
{\sl independent} of both temperature and luminosity.

We have computed bolometric magnitudes for the stars in our sample using the
(V-I) colours and the relation
\begin{displaymath}
BC_I \ = \ 0.02 \ + \ 0.575 (V-I) \ - \ 0.155 (V-I)^2
\end{displaymath}
derived from Leggett et al's (1996) data for field M dwarfs. Table 7 lists those
magnitudes, together with data for Hyades stars from Stauffer et al (1997), Jones et al (1996)
and Terndrup et al (2000). Deriving L$_\alpha$ for these stars requires an estimate of the
continuum flux at 6563\AA, F$_C$. Reid et al (1995b) demonstrated that F$_C$ can be derived from 
the R$_C$-band flux, F$_R$, from the relation
\begin{displaymath}
F_C \ = \ 1.45 F_R
\end{displaymath}
All of the stars in Table 1 have R-band photometry (Reid, 1993 and references therein). For
stars with multiple observations, the
H$\alpha$ fluxes are derived using the average equivalent width (with the exception
of the MJD 794.46 observation of RHy 281, discussed further below).
Variability, as noted above, leads to typical uncertainties of $\pm 15\%$.

The uppermost panel in figure 8 shows the (M$_{bol}$, log(L$_\alpha$/L$_{bol}$)) 
distribution for Hyades M dwarfs. Contrary to the suggestion by Terndrup et al (2000), 
There is no evidence for a decline in activity with decreasing 
luminosity. Known binaries and single stars are
distinguished using different symbols on the diagram, but there is no evidence that 
the former are more (or less) active than the latter. These data confirm results
derived previously by Reid et al (1995b), a paper which unaccountably seems to have
escaped Terndrup et al's attention.

Figure 8b plots the ($v \sin(i)$, M$_{bol}$) distribution for the
Hyades stars, illustrating the well known tendency towards higher rotation with
decreasing mass. However, figure 8c shows that there is no 
correlation between the level of chromospheric activity and rotation: a
Hyades M dwarf with $L_\alpha / L_{bol} \sim 1.5 \times 10^{-4}$ can have a measured
rotational velocity anywhere between $<2$ and 40 kms$^{-1}$, the full range
measured for cluster stars. 

This result is in good agreement with
data for late-type field M dwarfs (Hawley et al, 1999), but
directly contradicts Terndrup et al's analysis of a subset of the present sample.
The origin of the discrepancy rests with Terndrup et al's use of equivalent width 
as an estimator of stellar activity. As noted above (and, indeed, by 
Terndrup et al), the equivalent 
width of an emission line depends on both the total line flux and the level of
the local continuum. Thus, early-type Hyades M dwarfs have weaker emission than late-type dwarfs,
even though the chromosphere is at the same {\sl proportional} level of activity.
Further, figure 8b shows that late-type dwarfs (with higher H$\alpha$ EW) tend to
have more rapid rotation. Thus, the correlation between ``activity'' and rotation 
claimed by Terndrup et al is actually the (mass/M$_{bol}$, $v \sin(i)$) correlation
illustrated in figure 8b.

Terndrup et al also present rotation and activity measurements for low-mass Pleiades stars. Figure 9 plots
their data, together with observations from Jones et al (1996) and Basri \& Marcy (1995), using
$L_\alpha/L_{bol}$ rather than H$\alpha$ EW \footnote{Note that using (V-I) instead of M$_{bol}$ in
these diagram does not change the overall morphology}. With an age of $\sim120$ Myrs, the Pleiades 
stars are more luminous (and hotter) at a given mass, and significantly more active. Thus, the 
mass range is $\sim 0.1 < {M \over M_\odot} < 0.6$, similar to the Hyades stars. While the
measured rotational velocities approach 100 kms$^{-1}$, there is no evidence for any correlation
between rotation and the level of chromospheric activity.

These results are consistent with Hawley et al's (1999)  proposal that the mechanism driving 
chromospheric activity changes as one moves from early-type M dwarfs to mid- and late-type dwarfs.
As convection becomes increasingly important and the radiative core shrinks, the shell dynamo weakens.
In contrast, the turbulent dynamo increases in strength, supplanting the shell dynamo
entirely in fully convective stars. As a result, rotation is irrelevant to the level of activity
present in dwarfs of spectral type M3 and later.

\subsection {The flare on RHy 281}

One observation in Table 5 stands out: on December 12, 1997, the measured equivalent
width of H$\alpha$ emission in RHy 281 is $\sim 23$\AA. As figure 10
shows, we happened to catch the star during a flare outburst. In addition
to H$\alpha$ emission, He I 6678\AA\ is present with an equivalent width of 1.15\AA\,
core emission is detectable in the K I 7665/7699 doublet, and we measure equivalent
widths of 0.85 and 0.35 \AA\ for the 8500 and 8663 \AA\ components of the
Ca II near-infrared triplet. 

Since our observations concentrated almost exclusively on stars within a narrow range
of absolute magnitude, $11.5 < M_V < 16$, we can derive an approximate estimate
of the frequency of flare outbursts in these stars. In total, we expended 
$\sim 67,000$ seconds integrating on M dwarfs in the course of
this project; the RHy 281 flare
was detected in an 1800 second exposure, and we assume that the flare persisted at
a detectable level for the duration of the exposure. Given that assumption, 
we deduce that low-mass Hyades dwarfs are in outburst for $\sim 2.5\%$ of
the time. This compares with an upper limit of $< 7 \%$ for the duty cycle
estimated for the M9 field dwarf, BRI0021 (Reid et al, 1999). 

\section {Summary and conclusions}

We have obtained high resolution spectroscopy of 51 low-mass dwarfs in the Hyades cluster.
At least three are confirmed as spectroscopic binaries: two double-lined systems, RHy 42 and 244, the latter 
previously identified as a binary in HST imaging; and  RHy 403,
a short-period single-lined system with an extremely low-mass secondary. Based on the mass
function for the last star and the absence of any detection of the secondary at 8000\AA, 
we infer mass limits of $0.07 \le {M \over M_\odot} \le 0.095$, making this
unseen companion a strong brown dwarf candidate.

Four other stars are possible binaries: RHy 158 shows strong indications of velocity
variability; RHy 281 lies $\sim6$ kms$^{-1}$ from the
expected velocity of Hyades members; RHy 221 and 377 are also offset in velocity,
although to smaller extent. Three further stars have unusual cross-correlation functions, 
although it is not clear whether these stem from binarity or rapid rotation.
Combining the new discoveries with pevious HST and ground-based identifications, we
derive a binary fraction for Hyades M dwarfs of 23 to 30 \%, similar to 
results derived from nearby field M dwarfs.

We have also used our data to determine rotational velocities and explore possible
correlations between chromospheric activity and rotation. Combined with literature
data, we find that the level of chromospheric activity, as measured by the fractional
luminosity emitted in the H$\alpha$ line, $L_\alpha / L_{bol}$, is nearly constant for M
dwarfs with $8 < M_{bol} < 12$. There is no evidence for a significant correlation between
activity and rotation as measured by $v \sin(i)$. We interpret this as confirmation 
that the rotational dynamo is not present in fully convectice mid- and late-type
M dwarfs. Activity is probably driven by a shear dynamo mechanism, as
proposed by Hawley et al (1999). 

\subsection*{Acknowledgements}

Funding for SM was provided by STScI Grant GO-08146.01-97A.
INR would like to thank Wal Sargent for the allocation of Keck time in December, 1999.
This analysis is based on observations obtained at the W. M. Keck Observatory, which is operated
by the California Association for Research in Astronomy, a partnership of the University of California, 
the California Institute of Technology and the National Aeronautics and Space Administration.
The observatory was made possible by a generous grant from the W. M. Keck Foundation.


\begin{thebibliography}{}
\bibitem[Basri et al, 1995] {bas} Basri, G., Marcy, G.W. 1995, AJ, 109, 762
\bibitem[Burrows et al, 1993]{B93} Burrows, A., Hubbard, W.B., Saumon, D., Lunine, J.I. 1993,
ApJ, 406, 158
\bibitem[Cochran \& Hatzes, 1999]{ch99} Cochran, W.D., Hatzes, A.P. 1999, DPS, 31, 0902
\bibitem[Delfosse et al, 1998] {del98} Delfosse, X., Forveille, T., Perrier, C.,
Mayor, M. 1998, A\& A, 331, 581
\bibitem [Duquennoy \& Mayor] {dm91} Duquennoy, A. \& Mayor, M.  1991, A\&A, 248, 485
\bibitem[Durbey et al]{d93} Durney, B.R., De Young, D.S., Roxburgh, I. 1993 Sloar Physics, 145, 207 
\bibitem [Fischer \& Marcy (1992)] {fm92} Fischer, D.A. \& Marcy, G.W. 1992, ApJ, 396, 178
\bibitem[Giclas et al]{g72} Giclas, H.L., Burnham, R., Thomas, N.C., 1972, The Lowell 
Observatory Proper Motion Survey, Lowell Obs. Bull. 158
\bibitem[Gizis \& Reid]{gr1} Gizis, J.E., Reid, I.N. 1995, AJ, 110, 1248
\bibitem[Gray, 1992] {g92} Gray, D.F., 1992, in The Observation and Analaysis of Stellar
Photospheres (New York: Cambridge univ. Press)
\bibitem [Griffin {\sl et al.} (1988)] {g88} Griffin, R.F., Gunn, J.E., Zimmerman, B.A. \& Griffin, R.E.M.
1988, AJ, 96, 172
\bibitem[Hamburg]{h95} Hagen, H.-J., Groote, D., Engels, D., Reimers, D. 1995, A \& AS, 111, 195
\bibitem[Hanson, 1975] {h75} Hanson, R.B. 1975, AJ, 80, 379
\bibitem[Hawley et al, 1999] {h99} Hawley, S.L., Reid, I.N., Gizis, J.E. 1999, in {Workshop on Cool Stars and
Giant Planets}, {ed. C. Griffiths \& M. Marley}, ASP Conf. Ser. No. 214,
\bibitem[Henry, T.J] {h90} Henry, T.J. 1990, Ph.d. thesis, University of Arizona 
\bibitem[Henry \& McCarthy, 1993] {hm93} Henry, T.J., McCarthy, D.W. 1993, AJ, 106, 773
\bibitem[Jones et al, 1996]{j96} Jones, B.F., Fischer, D.A., Stauffer, J.R. 1996, AJ, 112, 1562
\bibitem [Leggett, 1992] {l92} Leggett, D.K., 1992, ApJS, 82, 351
\bibitem [Leggett {\sl et al.} (1994)] {ldh94} Leggett, S.K., Harris, H.C. \& Dahn, C.C. 1994, AJ, 108, 944
\bibitem[Leggett {\sl et al} (1996)]{l96} Leggett, S.K., Allard, F., Berriman, G., Dahn, C.C.,
Hauschildt, P. 1996, ApJS, 104, 117 
\bibitem[Marcy \& Benitz, 1989] {mb89} Marcy, G.W., Benitz, K.J. 1989, ApJ, 344, 441
\bibitem[Marcy \& Butler, 1992]{mb92} Marcy, G.W., Butler, R.P. 1992, PASP, 104, 270
\bibitem[Parker,  1955] {p55} Parker, E.N., 1955, ApJ, 122, 293
\bibitem[Patience et al] {pat} Patience, J., Ghez, A.M., Reid, I.N., Weinberger, A.J., Matthews, K.
1998, AJ, 115, 1972
\bibitem[Perryman et al, 1998]{p98} Perryman MAC, Brown AGA, Lebreton Y, G\'omez A, Turon C, Cayrel de Strobel G,
Mermilliod J-C, Robichon N, Kovalevsky J, Crifo F 1998. A\&A, 331, 81
\bibitem[Reid, 1992] {r92} Reid, I.N. 1992, MNRAS, 257, 257
\bibitem[Reid, 1993] {r93} Reid, I.N. 1993, MNRAS, 265, 785 
\bibitem [Reid et al, 1995] {rhg} Reid, I.N., Hawley, S.L., Gizis, J.E. 1995a, AJ, 110, 1838
\bibitem [Reid et al, 1995] {rhg} Reid, I.N., Hawley, S.L., Mateo, M. 1995a, MNRAS, 272, 828
\bibitem [Reid \& Gizis (1997a)] {rg97} Reid, I.N., Gizis, J.E. 1997a, AJ,  113, 2246
\bibitem [Reid \& Gizis (1997b)] {rg97b} Reid, I.N., Gizis, J.E. 1997b, AJ,  114, 1992
\bibitem[Reid \& Hawley, 199]{rh99} Reid, I.N., Hawley, S.L. 1999, AJ, 117, 343
\bibitem[Reid et al, 1999] {r99} Reid, I.N., Kirkpatrick, J.D., Gizis, J.E., Liebert, J. 1999, 
ApJ, 527, L105
\bibitem[Sandage \& Kowal]{sk86} Sandage, A., Kowal, C. 1986, AJ, 91, 1140
\bibitem[Skuamnich]{s72} Skumanich, A., 1972, ApJ, 171, 565
\bibitem [Stauffer et al, 1997] {s97} Stauffer, J.R., Balachandran, A.C., Krishnamurthi, A.,
Pinsonneault, M., Terndrup, D.M., Stern, R.A. 1997, ApJ, 475, 604
\bibitem[Stern et al, 1994] {st94} Stern, R. A., Schmitt, J.H.M.M, Pye, J.P., Hodgkin, S.T.,
Stauffer, J.R., Simon, T. 1994, ApJ, 427, 808
\bibitem[Terndrup et al, 2000]{t20} Terndrup, D.M., Stauffer, J.R., Pinsonneault, M.H., Sills, A.,
Yuan, Y., Jones, B.F., Fiscer, D., Krishnamurthi, A. 2000, AJ, 119, 1303
\bibitem[Tonry \& Davis]{td79} Tonry, J., Davis, M. 1979, AJ, 84, 1511
\bibitem[]{} Vogt, S.S. et al., 1994, S.P.I.E.,  2198, 362.

\end{thebibliography}
\end{document}